%% file: mp03-j.tex
\documentstyle[version,11pt,epsf,epsfig,fullpage]{article}

\input{psfig.sty}

\def\squareforqed{\hbox{\rlap{$\sqcap$}$\sqcup$}}
\def\qed{\ifmmode\squareforqed\else{\unskip\nobreak\hfil
\penalty50\hskip1em\null\nobreak\hfil\squareforqed
\parfillskip=0pt\finalhyphendemerits=0\endgraf}\fi}

\newcommand{\Remove}[1]{}
 \newcommand{\RemoveICALP}[1]{#1}
\newcommand{\IfICALP}[1]{}

\newcommand{\LineSpace}{1.1}
\renewcommand{\baselinestretch}{\LineSpace}\large\normalsize

 \include{epsf}

\def\papernumber #1 raised #2 {
  \vspace{-#2}
  \vbox to 0pt{\hfill\framebox{\bf Paper Number #1}}
  \vspace{#2}
}

\excludeversion{EXCLUDE}{}

\excludeversion{JOURNAL}{} 	
\newif\ifJournal      
\Journalfalse

\newif\ifFull   
\Fullfalse

\let\OldComment=\COMMENT
\def\COMMENT{\OldComment{$\clubsuit$}\footnotesize\sf}

\newtheorem{definition}{Definition}[section] 
\newtheorem{theorem}{Theorem}[section]
\newtheorem{lemma}[theorem]{Lemma}

\newtheorem{claim}[theorem]{Claim}
\newtheorem{corollary}[theorem]{Corollary}
\newtheorem{fact}[theorem]{Fact}

\input{macros}

\begin{document}

\excludeversion{EXCLUDE}
\includeversion{TM}
\includeversion{COMMENT}
\let\OldComment=\COMMENT
\def\COMMENT{\OldComment{$\clubsuit$}\footnotesize\tt}


\title{\Large\bf Efficient pebbling for list traversal synopses \footnote{%
A preliminary version of the results in this paper will be presented at ICALP'03}
}
\author{
\begin{tabular}{cc}
Yossi Matias\thanks{School of Computer Science, Tel Aviv University; {\tt matias@cs.tau.ac.il}.
	Research supported in part by the Israel Science Foundation.}
& Ely Porat\thanks{Department of Mathematics and Computer Science,
Bar-Ilan University, 52900 Ramat-Gan, Israel, (972-3)531-8407;
{\tt porately@cs.biu.ac.il}.}
\\
{\small Tel Aviv University} & {\small Bar-Ilan University}\\
&{\small  \& Tel Aviv University}\\
\end{tabular}
}

\date{}

\maketitle

 \thispagestyle{empty}
\begin{abstract}

We show how to support efficient back traversal in a unidirectional list,
 using small memory and with essentially no slowdown in forward steps.
Using $O(\log n)$ memory for a list of size $n$, the $i$'th back-step from the
 farthest point reached so far takes $O(\log i)$ time in the worst case, while the
 overhead per forward step is at most $\epsilon$ for arbitrary small constant
 $\epsilon>0$.
An arbitrary sequence of forward and back steps is allowed.
A full trade-off between memory usage and time per back-step is presented: $k$ vs.\ $kn^{1/k}$
 and vice versa.
Our algorithms are based on a novel pebbling technique which moves pebbles
 on a virtual binary, or $t$-ary, tree that can only be traversed in a pre-order fashion.

The compact data structures used by the pebbling algorithms, called list traversal
 synopses, extend to general directed graphs, and have
 other interesting applications, including memory efficient hash-chain
 implementation.
Perhaps the most surprising application is in showing that for any program,
 arbitrary rollback steps can be efficiently supported with small
 overhead in memory, and marginal overhead in its ordinary execution.
More concretely: Let $P$ be a program that runs for at most $T$ steps,
 using memory of size $M$. Then, at the cost of recording the
 input used by the program, and increasing the memory by a factor of
$O(\log T)$ to $O(M \log T)$, the program $P$ can be extended to support an
 arbitrary sequence of forward execution and rollback steps:
 the $i$'th rollback step takes $O(\log i)$ time in the worst case, while
 forward steps take $O(1)$ time in the worst case, and $1+\epsilon$
 amortized time per step.


\end{abstract}

\setlength{\parindent}{0.0in}
\setlength{\parskip}{0.1 in}
\section{Introduction\label{sec:intro}}
A unidirectional list enables easy forward traversal in constant time per step. 
However, getting from a certain object to its preceding object cannot be done effectively.
It requires forward traversal from the beginning of the list and takes time proportional to the
distance to the current object, using $O(1)$ additional memory.
In order to support more effective back-steps on a unidirectional list,
it is required to add auxiliary data structures. 

Trailing pointers, consisting of a backward pointers from the  
current position to the beginning of the list,
can be easily maintained in $O(1)$ time per forward step, and support back-steps in $O(1)$
time. However, the memory required for maintaining trailing pointers is $\Theta(n)$,
where $n$ is the distance from the beginning of the list to the farthest point reached
so far. A simple time-memory trade-off can be obtained by keeping a pointer every $n/k$
forward steps. With memory of size $\Theta(k)$, each back-step can be done in $\Theta(n/k)$ 
time. This provides a full generalization of the two previous solutions,
with $\Theta(n)$ memory-time product.

A substantially better trade-off can be obtained, using what we call {\em skeleton} 
 data structures.
These skeletons enable to obtain full back traversals in $O(k n^{1/k})$ amortized 
 time per back-step, using $k$ additional pointers~\cite{BP99}.
However, if one wishes to support fully dynamic list traversal consisting of an
arbitrary sequence of forward and back steps, then managing the pointers positions
becomes challenging. For the further restriction that forward steps do not incur
more than constant overhead (independent of $k$), the problem becomes even more 
difficult.

The goal of this work is to support memory- and time-efficient back traversal in
unidirectional lists, without essentially increasing the time per forward traversal.
In particular, under the constraint that forward steps should remain constant, we
would like to minimize the number of pointers kept for the lists, 
the memory used by the algorithm, and the time per back-step,
supporting an arbitrary sequence of forward and back steps.

Of particular interest are situations in which the unidirectional list is already given, 
and we have access to the list but no control over its implementation. 
The list may represent a data structure
implemented in computer memory or in a database, or it may reside on a separate computer system.
The list may also represent a computational process, where the objects in the list are 
configurations in the computation and the next pointer represents a computational step.
Supporting efficient back traversal on the list enables effective program rollback, and
 requiring $O(1)$ time per forward step implies that forward execution of the program is 
 not significantly affected.

To address such variety of scenarios more accurately, we may assume that the lists could 
 be accessed via a third party, denoted as the PSP (for Pointer Service Provider). 
The traversal algorithm communicates to the PSP only instructions of type 
 {\it forward, fetch, free,} and {\it create,} with pointers identifications. 
The complexity metric accounts separately for the pointers kept at the PSP, which will be 
 represented  throughout the paper as {\em pebbles}, and the data 
 structure used by the traversal algorithm. 
The forward steps requested from the PSP are counted separately, and are denoted 
 as {\it list-steps}.

\subsection{Main results}
The main result of this paper is an algorithm that supports efficient back traversal in 
 a unidirectional list, using small memory and with essentially no slowdown in forward 
 steps: $1+\epsilon$ 
 amortized
 time per forward step for arbitrary small constant $\epsilon >0$,
 and $O(1)$ time in the worst case.
Using $O(\log n)$ memory, back traversals can be supported in $O(\log n)$ time 
per back-step, where $n$ is the distance from the beginning of the list to farthest point
reached so far. 
%
%
In fact, we show that a back traversal of limited scope can be executed 
more effectively: $O(\log i)$ time for the $i$'th back-step, for any $i \le n$,
using 
$O(\log n)$ memory. 

More generally, the following trade-offs are obtained: 
$O(k n^{1/k})$ time per back-step, using $k$ additional pointers,
or $O(k)$ time per back-step, using $O(k n^{1/k})$ additional pointers; in 
both cases supporting $O(1)$ time per forward step (independent of $k$).
Our results extend to general directed graphs, with additional memory
of $\log d_v$ bits for each node $v$ along the backtrack path, where $d_v$ 
is the outdegree of node~$v$.

For the PSP model, our main result is a {\em list pebbling algorithm} that uses $\log n$ 
 pebbles and $O(\log n)$ memory to support the $i$'th back-step in $O(\log i)$ list-steps and 
 $O(\log i)$ time, with $\epsilon$ 
 amortized
 overhead per forward step, for arbitrary small constant $\epsilon >0$.

The crux of the list traversal algorithm 
is an efficient pebbling technique which moves 
pebbles on virtual binary or $t$-ary trees that can only be traversed in a pre-order fashion.
We introduce the {\em virtual pre-order tree} data structure which enables managing the
pebbles positions in a concise and simple manner, and the {\em recycling bin} 
 data structure that manages pebbles allocation.

\subsection{Applications}
Consider a program $P$ running running in time $T$.
Then, using our list pebbling algorithm, the program can be extended to a program $P'$ 
 that supports rollback steps, where a rollback after step $i$ means that the program 
 returns to the configuration it had after step $i-1$. 
Arbitrary ad-hoc rollback steps can be added to the execution of the program $P'$ at a cost of 
 increasing the memory requirement by a factor of $O(\log T)$, and having the $i$'th 
 rollback step supported in $O(\log i)$ time. The overhead for the forward execution
 of the program can be kept an arbitrary small constant.

Allowing effective rollback steps may have interesting applications. For instance,
a desired functionality for debuggers is to allow pause and rollback during execution.
Another implication is the ability to take programs that simulate processes and allow
running them backward in arbitrary positions. 
Thus a program can be run with $\epsilon$ overhead in its normal execution, and allow 
 pausing at arbitrary points, and running backward an arbitrary number of steps with 
 logarithmic time overhead per back-step. 
The memory required is keeping state 
configuration of $\log T$ points, and additional $O(\log T)$ memory.
Often, debuggers and related applications avoid keeping full program states by
keeping only differences between the program states. If this is allowed, then a more
appropriate representation of the program would be a linked list in which every node
represents a sequence of program states, such that the accumulated size of the differences 
is in the order of a single program state.

Our pebbling technique can be used to support backward computation of a hash-chain in time
 $O(k n^{1/k})$ using $k$ hash values, or in time $O(k)$ using $O(k n^{1/k})$ hash values, 
 for any $2\le k \le \log n$.
A {\em hash-chain} is obtained by repeatedly applying a one-way hash function, starting 
with a secret seed. 
There are a number cryptographic applications, 
 including password authentication~\cite{lamport81},
 micro-payments~\cite{Rivest:Shamir:96}, forward-secure 
signatures~\cite{Itkiss-Reyzin,Kozlov-Reyzin},
 and broadcast authentication protocol~\cite{Perrig}.
Our results enable effective implementation with arbitrary memory size.

The list pebbling algorithm extends to directed trees and general directed graphs. 
Applications include the effective implementation of the parent function (``..'') 
for XML trees, and effective graph traversals with applications to ``light-weight'' 
Web crawling and garbage collection. 

\subsection{Related work}


If it is allowed to change pointers in the list, then one can use the Schorr-Waite 
 algorithm~\cite{SW67}. 
This algorithm enables constant time back-step by simply utilizing the ``next'' pointers
at the nodes from the head of the list to the current position to hold pointers to the 
previous nodes. Constant size auxiliary memory is sufficient to support this ``in place''
algorithm. The Schorr-Waite algorithm also works for trees, dags, and general directed 
graphs. 

Another solution for an in-place encoding which supports back traversal is based 
on the following technique. For each node $v$, instead of keeping the pointer next($v$), 
we keep the XOR of prev($v$) and next($v$). As in the Schorr-Waite algorithm,
only two auxiliary pointers are required -- for the current position and for the previous
position. At any position $v$, moving forward and backward can be done in constant time
using the encoded information. This algorithm has the advantage that the list encoding 
remains intact during traversal, and unlike for the Schorr-Waite algorithm, multiple 
users can traverse the list. It can also be extended to trees, dags, and general 
directed graphs with similar advantage. 

Recall that both algorithms do not fit the requirement that the list (trees, digraphs)
cannot be altered. In particular, these algorithms cannot be used for the applications 
in which the list represents a computation.

The Schorr-Waite algorithm~\cite{SW67} has numerous applications;
 see e.g.,~\cite{sobel99recycling,walker00alias,chung00reducing}.
It would be interesting to explore to what extent these applications could
benefit from the non-intrusive nature of our algorithm.
There is an extensive literature on graph traversal with bounded memory but for other
problems than the one addressed in this paper; see, e.g.,~\cite{hirschberg93boundedspace,bender98power}.
Pebbling models were extensively used for bounded space upper and lower
bounds. See e.g., the seminal paper by Pippenger~\cite{Pip82} and more 
recent papers such as~\cite{bender98power}.

The closest work to ours is the recent paper by Ben-Amram and Petersen~\cite{BP99}. 
They present a clever algorithm that, using memory of size $k\le \log n$, 
 supports back-step in $O(k n^{1/k})$ time. 
However, in their algorithm forward steps take $O(k)$ time. Thus, 
their algorithm supports $O(\log n)$ time per back-step, using $O(\log n)$ memory but 
with $O(\log n)$ time per forward step, which is unsatisfactory in our context. 
Ben-Amram and Petersen also prove a near-matching lower bound, implying that to support 
 back traversal in $O(n^{1/k})$ time per back-step it is required to have $\Omega(k)$ 
 pebbles.
Our algorithm supports similar trade-off for back-steps as the Ben-Amram Petersen algorithm,
 while supporting simultaneously constant time per forward step. 
In addition, our algorithm extends to support $O(k)$ time per back-step, using memory
 of size $O(k n^{1/k})$, for every $k\le \log n$,

%

Recently, and independently to our work, Jakobsson and Coppersmith~\cite{Ja02,CJ02}
 proposed a so-called fractal-hashing technique that enables backtracking hash-chains in 
 $O(\log n)$ amortized time using $O(\log n)$ memory.  
Thus, by keeping $O(\log n)$ hash values along the hash-chain, their algorithms 
 enables, starting at the end of the chain, to get repeatedly the preceding hash value
 in $O(\log n)$ amortized time.
Subsequently, Sella~\cite{Sella03} showed how to generalize the fractal hashing scheme, to
 work with $k$ hash values, for any $k < \log n$, supporting back steps in $O(k n^{1/k})$.
These works are only in the context of hash chains and do not
deal with efficient forward traversal. Note also that our
pebbling algorithm enables a full memory-time trade-off
for hash-chain
 execution; that is, both $O(k)$ memory and $O(k n^{1/k})$ time per back step and vice versa.
In addition, it is guaranteed that the time per execution is bounded in the worst case. 

The most challenging aspect of our algorithm is the proper management of the 
 pointers positions under the restriction that forward steps have very little 
 effect on their movement, to achieve $\epsilon$-overhead per forward step. 
This is obtained by using the virtual pre-order tree data structure in conjunction 
 with a so-called recycling-bin data structure and other techniques, 
to manage the positions of the back-pointers in a concise and simple manner.

\subsection{Outline}
The rest of the paper is organized as follows. 
In Section~\ref{Section:Skeleton-and-tree} 
we describe the skeleton data structures that provide some intuition about
 the pebbling techniques, and the virtual pre-order tree data structure, which will be 
 used by all our algorithms. 
In Section 3 we describe the full algorithm, called the list pebbling algorithm, which
 obtains $O(\log n)$ amortized time per back-step using $O(\log n)$ pebbles, while 
 supporting $O(1)$ time per forward step.
The advanced list pebbling algorithm, is described in Section~4. This algorithm supports 
 $O(\log n)$ time per back-step in the worst case, using $\log n$ pebbles, as well as
 supporting $\epsilon$ overhead per forward step. 
An extension of the algorithm to support full time-memory trade-off of 
 $O(k)$ vs.\ $O(k n^{1/k})$ is described in Section~5. 
In Section~6 we describe the application for efficient reversal of program execution, 
 and for efficient processing of hash-chains. 
Extensions to trees and other graphs are given in Section~7, and we conclude in
 Section~8.
Earlier versions of this paper appear in~\cite{MP02,MP03-icalp}.

\Section{The skeleton and virtual pre-order tree data structures}{Skeleton-and-tree}
%


In this section we illustrate the basic idea of the list pebbling algorithm, 
and demonstrate it through a limited functionality of having a sequence of back-steps only. 

We first describe in Section~\ref{Section:Skeleton}
algorithms based on {\it skeleton} data structures, of which the
most advanced supports a sequence of back-steps in $O(\log n)$ amortized time per back-step, 
using $\log n$ pebbles. 
These data structures are similar in nature to the ones used by~\cite{BP99,%
Ja02,CJ02,sliding}.

A full algorithm must support an arbitrary sequence of forward and backward steps,
and we will also be interested in refinements, such as reducing to minimum the number 
of pebbles. 
Adapting the skeleton data structures to support the full algorithm and its 
refinements may be quite complicated, since controlling and handling the positions of 
the various pointers becomes a challenge. 
For the further restriction that forward steps do not incur
more than constant overhead (independent of $k$), the problem becomes even more 
difficult and we are not aware of any previously known technique to handle this.

To have control over the pointers positioning, we present in
Section~\ref{Section:pre-order-tree} the {\it virtual pre-order tree} 
data structure, and show how it supports the sequence of back-steps similarly to the 
skeleton data structure. 
In the next sections, we will see how the virtual pre-order tree data 
structure is used to support the full algorithm as well as more advanced algorithms.

\Remove{
To have control over the pointers positioning, we present the {\it virtual pre-order tree} data 
structure, and show how it supports the sequence of back-steps similarly to the skeleton 
data structure. In the next sections, we will see how the virtual pre-order tree data 
structure is used to support the full algorithm as well as more advanced algorithms.

\Subsection{The skeleton data structure}{Skeleton}
Let $n$ be the current position and assume that $n$ is a power of $2$. We maintain 
$\log n$ pointers between the beginning of the list to the current position, where the 
$i$'th pointer is at distance $2^i$ from the current position, $i=0,1,\ldots,\log n$. 
Denote this as the skeleton data structure of size $n$.

A sequence of back traversals from position $n$ to the position of the $i$'th pointer 
is done as follows:
\begin{enumerate}
\item Have a sequence of back traversals from position n to the position of the $i-1^{st}$
pointer, using the skeleton data structure.
\item Build a skeleton data structure of size $2^{i-1}$ between the positions of the 
$i$'th and $i-1^{st}$ pointers, using the pointers freed in step 1. (See
 Figure~\ref{Fig:skeleton})
\item Have a sequence of back traversals from the position of the $i-1^{st}$ pointer
to the position of the $i$'th pointer, using the skeleton data structure between 
these points.
\end{enumerate}
Steps 1 and 3 are a recursive application of the algorithm for problems of size $2^{i-1}$.
The time required to have a sequence of back-steps from position $n$ to the position of
the $i$'th pointer is therefore $T(i) = 2T(i-1) + 2^{i-1}$, $T(1)= O(1)$, implying 
$T(i) = i 2^i$, and an amortized $O(\log d)$ time for the $d$'th back-step. 
}

\RemoveICALP{
\Subsection{The skeleton data structure}{Skeleton}
In this subsection we illustrate simple skeleton data structures and demonstrate 
them through a limited functionality of having a sequence of back-steps only. 

As a motivating example, we outline first how to obtain $O(\sqrt{n})$ amortized time 
 per back-step, using two additional pointers, $p$ and $p'$. 
For simplicity, let us only describe how to implement a sequence of back-traversals 
 from position $n$ to the beginning of the list. 
When positioned at node $n$, pointer $p'$ acts as a {\em shadow pointer}, and points 
 to position $n-\sqrt{n}$. 
As long as the current position is between pointer $p'$ and position $n$, a back-step
 is implemented by advancing pointer $p$, which acts as an {\em assisting pointer}, 
 from position $p'$, until next($p$) becomes the current position. 
When the current position becomes $p'$, we will update $p'$ to be $n-2\sqrt{n}$ by moving 
 the assisting pointer $p$ 
 forward $n-2\sqrt{n}$ steps starting from the beginning of the list. 
Each update occurs only after $\sqrt{n}$ back steps are executed since the previous 
 update. 
Therefore, the amortized cost per back-step is smaller than $\sqrt{n}$. 
It is straightforward to extend this into a back traversal all the way to the 
 beginning of the list.

An improved, $O(n^{1/3})$ amortized time per back step can be obtained by having the 
 shadow pointer $p'$ positioned at location $n-n^{2/3}$, and adding
 a second shadow pointer $p''$, pointing initially to position $n-n^{1/3}$. 
As long as the current position is between pointer $p''$ and position $n$, a back-step
 is implemented by advancing the assisting pointer $p$, from position $p''$, 
 $O(n^{1/3})$ steps until 
 next($p$) becomes the current position. 
When the current position becomes $p''$, we will update $p''$ to be $n-2n^{1/3}$ 
 by moving forward $n^{2/3}-2n^{1/3}$ steps starting from position $p'$ (after
 $n^{1/3}$ steps each taking $O(n^{1/3})$ time). 
When the position of $p''$ becomes $p'$ (after $n^{2/3}$ steps each taking
 $O(n^{1/3})$ time), we will update $p'$ to be $n-2n^{2/3}$, 
 by moving forward $n-2n^{2/3}$ steps starting from the beginning of the list. 
This results with at most $3 n^{1/3}$ amortized time per back step.

Using a related technique, a full back-traversal can be implemented, in which each
back-step takes $O(1)$ amortized time, using $O(\sqrt{n})$ additional pointers.
\RemoveICALP{
When positioned at node $n$, we keep $\sqrt{n}$ shadow pointers at positions 
 $n-\sqrt{n}+1$ through $n$, as well as $\sqrt{n}$ pointers at positions $n-i\sqrt{n}$, 
 for $i=1,\ldots,\sqrt{n}-1$.
As long as the current position is between $n-\sqrt{n}+1$ and $n$, each back-step 
 takes $O(1)$ time. 
When the current position reaches position $n-\sqrt{n}$, the $\sqrt{n}$ shadow pointers
 are moved to positions $n-2\sqrt{n}+1$ through $n-\sqrt{n}$, in $\sqrt{n}$ time, 
 or $O(1)$ amortized time per back-step.
}

These methods can be extended to more generally support full back traversals 
in $O(k n^{1/k})$ amortized time per back-step, using $k$ additional pointers,
or in $O(k)$ amortized time per back-step,  using $O(k n^{1/k})$ additional pointers.


\Remove{
However, if one wishes to support fully dynamic list traversal consisting of an
arbitrary sequence of forward and back steps, then managing the pointers positions
becomes challenging. For the further restriction that forward steps do not incur
more than constant overhead (independent of $k$), the problem becomes even more 
difficult and we are not aware of any previously known technique to handle this.

To have control over the pointers positioning, we present the {\it virtual pre-order tree} 
data structure, and show how it supports the sequence of back-steps similarly to the 
skeleton data structure. 
In the next sections, we will see how the virtual pre-order tree data 
structure is used to support the full algorithm as well as more advanced algorithms.
}

The skeleton data structure, described next, supports a sequence of 
back-steps only in $O(\log n)$ amortized time per back-step, using $\log n$ pebbles. 

Let $n$ be the current position and assume that $n$ is a power of $2$. 
We maintain $\log n$ pointers between the current position and the beginning of the 
 list, where the $i$'th pointer is at distance $2^i$ from the current position, 
 $i=0,1,\ldots,\log n$. 
Denote this as the {\em skeleton} data structure of size $n$.

A sequence of back traversals from position $n$ to the position of the $i$'th pointer 
is done as follows:
\begin{enumerate}
\item Have a sequence of back traversals from position n to the position of the $i-1^{st}$
pointer, using the skeleton data structure.
\item Build a skeleton data structure of size $2^{i-1}$ between the positions of the 
$i$'th and $i-1^{st}$ pointers, using the pointers freed in step 1. (See
 Figure~\ref{fig:skeleton}).
\item Have a sequence of back traversals from the position of the $i-1^{st}$ pointer
to the position of the $i$'th pointer, using the skeleton data structure between 
these points.
\end{enumerate}
Steps 1 and 3 are recursive applications of the algorithm for problems of size $2^{i-1}$.
Step 2 is implemented in a single sequence $2^{i-1}$ forward steps.
The time required to have a sequence of back-steps from position $n$ to the position of
the $i$'th pointer is therefore $T(i) = 2T(i-1) + 2^{i-1}$, $T(1)= O(1)$, implying 
$T(i) = O(i 2^i)$, and an amortized $O(\log d)$ time for the $d$'th back-step. 


\Remove{
Recall that a full algorithm must support an arbitrary sequence of forward and backward steps,
and we will also be interested in refinements, such as reducing to minimum the number 
of pebbles. Adapting the skeleton data structure to support the full algorithm and its 
refinements may be quite complicated, since controlling and handling the positions of 
the various pointers becomes a challenge. 
}

\begin{figure}[htb]
\centerline{\psfig{figure=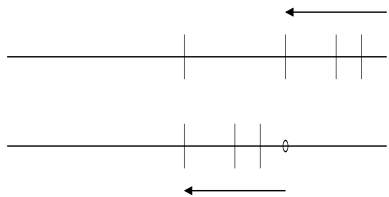,height=1.0in,width=2.0in}}
\caption{The skeleton data structure}
\label{fig:skeleton}
\end{figure}

}

\begin{figure}[htb]
\centerline{\psfig{figure=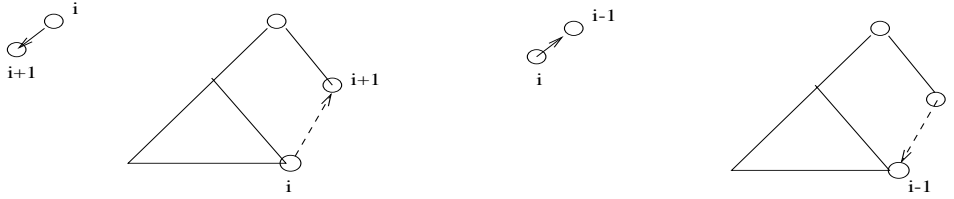,height=1.0in,width=5.0in}}
\caption{Preorder traversal}
\label{fig:pre}
\end{figure}

\Subsection{The virtual pre-order tree data structure}{pre-order-tree}
The reader is reminded (see Figure~\ref{fig:pre}) that in a pre-order traversal, 
 the successor of an internal node in the tree is always its left child; 
 the successor of a leaf that is a left child is its right sibling; 
 and the successor of a leaf that is a right child is defined 
 as the right sibling of the nearest ancestor that is a left child.
An alternative description is 
 as follows: 
 consider the largest sub-tree of which this leaf is the right-most leaf, 
 and let $u$ be the root of that sub-tree. 
Then the successor is the right-sibling of $u$.
Consequently, the backward traversal on the tree will be defined as follows. 
The successor of a node that is a left child is its parent. The successor of a node $v$
that is a right child is the rightmost leaf of the left sub-tree of $v$'s parent. 

The {\it virtual pre-order tree} data structure consists of (1) an implicit binary tree, 
whose nodes correspond to the nodes of the linked list, in a pre-order fashion, 
and (2) an explicit sub-tree of the implicit tree, whose nodes are pebbled. 
For the basic algorithm, the pebbled sub-tree consists of the path from the root to the 
current position.

Each pebble represents a pointer; i.e., pebbled nodes can be accessed in constant time.
We defer to later sections the issues of how to maintain the pebbles, and how to navigate
within the implicit tree, without actually keeping it. 
\begin{figure}[htb]
\centerline{\psfig{figure=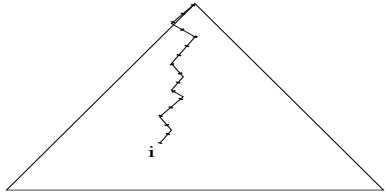,height=1.0in,width=2.0in}}
\caption{A pebbled path from the root to current position i}
\label{fig:path}
\end{figure}
\RemoveICALP{
\Subsection{Back traversal using the virtual pre-order tree data structure}{back-traversal}
}
Starting at node $n$, a back-traversal can be executed while maintaining $\log n$ pebbles
with $O(\log n)$ amortized time per back-step, as follows.

If node $i$ is a left child, then node $i-1$ is the parent of $i$, and the path from the 
root to node $i-1$ is already pebbled. Therefore, doing the backtrack step as well as 
updating the data structure are trivial. 

If node $i$ is a right child, then node $i-1$ is the rightmost leaf in the sub-tree, $T'$,
whose root is the left sibling of node $i$. In this case the path from the root of $T'$ 
to node $i-1$ (consisting of going down $T'$ through right children only), is yet to be 
pebbled. The challenge is that getting into these nodes requires a full traversal of $T'$.

Let $T$ be the sub-tree whose root is node $i$, and let $t$ be the size of $T$. 
Note that the size of $T'$ is also $t$. 
Thus, moving from node $i$ to node $i-1$, as well as pebbling the path from the 
 root of $T'$ to node $i-1$ takes $t$ steps.
We will charge this cost to the sequence of all backtrack steps within $T$, i.e., 
 starting from the rightmost leaf in $T$ and getting to its root (node $i$).

The total cost of all backtrack steps is $C= \sum_v t(v)$, where $v$ is a right child and 
$t(v)$ is the size of the subtree rooted at $v$. It is easy to verify that 
$C <$$n \log n \over 2$, resulting with amortized $O(\log n)$ time per back-step. 
In fact, it is not difficult to show that for every prefix of size n' 
 the amortized time per back-step is $O(\log n')$.

\Section{The list pebbling algorithm}{List-pebbling-algorithm}
In this section we describe the list pebbling algorithm, which supports an arbitrary 
 sequence of forward and back steps. 
Each forward step takes $O(1)$ time, where each back-step takes $O(\log n)$ amortized 
 time, using $O(\log n)$ pebbles. 
We will first present the basic algorithm which uses $O(\log^2 n)$ pebbles, then 
 describe the pebbling algorithm which uses $O(\log n)$ pebbles without considerations 
 such as pebble maintenance, and finally describe a full implementation using a 
 so-called recycling bin data structure.

The list pebbling algorithm is an extension of the algorithm described in Section 2.3.
It uses a new set of pebbles, denoted as {\it green pebbles}. The pebbles used as described 
in Section 2 are now called {\it blue pebbles}. The purpose of the green pebbles is to 
be kept as placeholders behind the blue pebbles, as those are moved to new nodes in 
forward traversal. Thus, getting back into a position for which a green pebble is 
still in place takes $O(1)$ time.

\begin{figure}[htb]
 \centerline{\psfig{figure=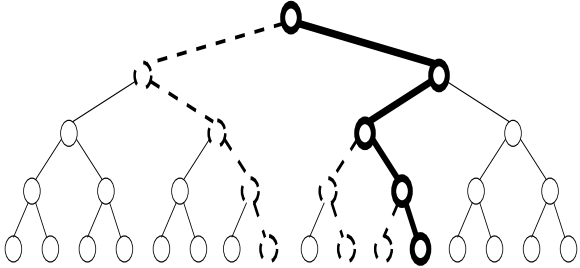,height=1.4in,width=3.0in}}
\caption{Green subpaths (dashed lines)}
\label{fig:green}
\end{figure}

\Subsection{The basic list pebbling algorithm}{basic-algorithm}
Define a {\it left-subpath (right-subpath)} as a path consisting of nodes that are all 
left children (right children). Consider the (blue-pebbled) path $p$ from the root to 
node $i$. We say that $v$ is a left-child of $p$ if it has a right sibling that is in $p$ 
(that is, $v$ is not in $p$, it is a left child, and its parent is in $p$ but not the node
i). As we move forward, green pebbles are placed on right-subpaths that begin at left 
children of $p$ (see Figure~\ref{fig:green}).
Since $p$ consists of at most $\log n$ nodes, the number of green 
pebbles is at most $\log^2 n$. 

When moving backward, green pebbles will become blue, and as a result, their left 
subpaths will not be pebbled. Re-pebbling these sub-paths will be done when needed.
When moving forward, if the current position is an internal node, then $p$ is extended 
with a new node, and a new blue pebble is created. No change occurs with the green 
pebbles. If the current position is a leaf, then the pebbles at the entire right-subpath 
ending with that leaf are converted from blue to green. Consequently, all the green 
sub-paths that are connected to this right-subpath are un-pebbled. That is, their 
pebbles are released and can be used for new blue pebbles.

We consider three types of back-steps:\\
{\it (i) Current position is a left child:}
The predecessor is the parent, which is on $p$, and hence pebbled. Moving takes $O(1)$ time;
 current position is to be un-pebbled.\\
{\it (ii) Current position is a right child, and a green sub-path is connected to its parent:} 
Move to the leaf of the green sub-path in $O(1)$ time, convert the pebbles on this 
 sub-path to blue, and un-pebble the current position. \\
{\it (iii) Current position is a right child, and its parent's sub-path is not pebbled:}
Reconstruct the green pebbles on the right sub-path connected to its parent $v$, 
 and act as in the second case. 
This reconstruction is obtained by executing forward traversal of the left sub-tree of $v$. 
We amortize this cost against the sequence of back-steps starting at the right sibling 
 of $v$ and ending at the current position. 
This sequence includes all nodes in the right sub-tree of $v$. 
Hence, each back-step is charged with one reconstruction step in this sub-tree. 
\begin{Claim}{back-step}
Each back step can be charged at most $\log n$ times.
\end{Claim}
\begin{Proof}
Consider a back step from a node $u$. The claim follows from the fact that such back-step 
 can only be charged once for each complete sub-tree that~$u$ belongs to.
\end{Proof}
We can conclude:
\begin{Theorem}{basic-list-pebbling}
The basic list pebbling algorithm supports $O(\log n)$ amortized list-steps per 
back-step, one list-step per forward step, using $O(\log^2 n)$ pebbles.
\end{Theorem}

\subsection{The list pebbling algorithm with $O(\log n)$ pebbles}
\begin{figure}[htb]
\centerline{\psfig{figure=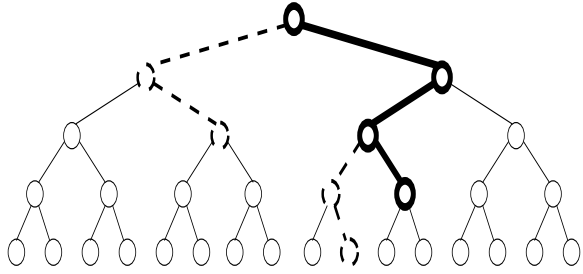,height=1.4in,width=3.0in}}
\caption{Using $\log n$ pebbles}
\label{fig:logn}
\end{figure}

The basic list pebbling algorithm is improved by the reducing the number of green
pebbles on most of the green paths.
Let $v$ be a left child of $p$ and let $v'$ be the right sibling of $v$. 
Denote $v$ to be the {\it last left child} of $p$ if the left subpath starting 
 at $v'$ ends at the current position; 
 let the right subpath starting at the last left child be the {\it last right subpath}. 
Then, if $v$ is not the last left child of $p$, the number of pebbled nodes in the 
 right subpath starting at $v$ is at all time at most the length of the left 
 subpath in $p$, starting at $v'$ (see Figure~\ref{fig:logn}). 
If $v$ is the last left child of $p$, the entire right subpath starting at $v$ can 
 be pebbled. 
We denote the (green) right subpath starting at $v$ as the {\it mirror subpath} of 
 the (blue) left subpath starting at $v'$. 
Nodes in the mirror subpath and the corresponding left subpath are said to be 
 {\it mirrored} according to their order in the subpaths. 
The following clearly holds: 

\begin{Claim}{green-pebbles-number}
The number of green pebbles is at most $\log n$.
\end{Claim}

A sequence of forward steps and corresponding blue path and green path of
each position is depicted in Figure~\ref{forward-traversal}

When moving forward, there are two cases:\\
(1) {\it Current position is an internal node:} as before, $p$ is extended with a new node, 
and a new blue pebble is created. No change occurs with the green pebbles (the 
mirror subpath begins at the last left child of $p$).\\
(2) {\it Current position i is a leaf that is on a right subpath starting at v (which could be i, if i is a left child):} 
we pebble (blue) the new position, which is the right sibling of $v$, and the pebbles 
at the entire right subpath ending at $i$ are converted from blue to green. Consequently,
(1) all the green sub-paths that are connected to the right subpath starting at v are 
un-pebbled; and (2) the left subpath in $p$ which ended at $v$ now ends at the parent of 
$v$, so the mirror (green) node to $v$ should now be un-pebbled. 
The released pebbles can be reused for new blue pebbles.

Moving backward is similar to the basic algorithm. There are three types of back-steps.\\
(1) {\it Current position is a left child:} predecessor is the parent, which is on $p$, and hence 
pebbled. Moving takes $O(1)$ time; current position is to be un-pebbled. No change occurs 
with green pebbles, since the last right subpath is unchanged.\\
(2) {\it Current position is a right child, and the (green) subpath connected to its parent is entirely pebbled:} 
Move to the leaf of the green subpath in $O(1)$ time, convert the pebbles on this 
subpath to blue, and un-pebble the current position. Since the new blue subpath is a 
left subpath, it does not have a mirror green subpath. However, if the subpath begins 
at $v$, then the left subpath in $p$ ending at $v$ is not extended, and its mirror green 
right subpath should be extended as well. This extension is deferred to the time the 
current position will become the end of this right subpath, addressed next. \\
(3) {\it Current position is a right child, and the (green) subpath connected to its parent is only partially pebbled:} 
Reconstruct the green pebbles on the right subpath connected to its parent $v$, and act 
as in the second case. This reconstruction is obtained by executing forward traversal of 
the sub-tree $T_1$ starting at $v$, where $v$ is the last pebbled node on the last 
right subpath. We amortize this cost against the back traversal starting at the right 
child of the mirror node of $v$ and ending at the current position. This sequence 
includes back-steps to all nodes in the left sub-tree $T_2$ of the mirror of $v$.
This amortization is valid since the right child of $v$ was un-pebbled in a forward 
step in which the new position was the right child of the mirror of $v$.
Since the size of $T_1$ is twice the size of $T_2$, each back-step is charged with at 
most two reconstruction steps.

As in \Ref{Claim}{back-step}, we have 
that each back step can be charged at most $\log n$ times, resulting with:

\begin{Theorem}{2logn-pebbles}
The list pebbling algorithm supports full traversal in at most $\log n$ amortized list-steps per 
back-step, one list-step per forward step, using $2\log n $ pebbles.
\end{Theorem}

\RemoveICALP{
\Subsection{A low-overhead, run-time sensitive implementation}{refined-analysis}
Using a more careful analysis and easy refinement of the list pebbling algorithm we can show 
the following.

\begin{Theorem}{epsilon-overhead}
The list pebbling algorithm supports $O(\log i)$ 
amortized time per back-step, 
$\epsilon$ amortized time overhead per forward step, using $O(\log n)$ 
pebbles, where $i$ is the distance from the current position to the farthest point 
traversed so far, and $\epsilon$ is an arbitrary small constant.
\end{Theorem}

\begin{Proof}
Let $T$ be the smallest tree that includes the farthest point $f$ traversed 
so far and the current position $i$. When traversing within $T$, no change occurs in 
nodes outside of T. The sequence of back-steps from $f$ to $i$ consists of two 
sub-sequences: from $f$ to $r+1$, which is in the right subtree $T_r$ of $T$, and from $r$
to $i$, which is in the left subtree $T_l$ of $T$. Note that $r$ is the rightmost leaf 
in $T_l$, and $r+1$ is the root of $T_r$. 
Similar to \Ref{Claim}{back-step}, note that each sub-step occurs at a node that is in 
 at most $\log i$ subtrees, and therefore each back-step can be charged at 
 most $\log i$ times.

To obtain $\epsilon$ overhead per forward step, we use the following simple method, 
which is applicable to all list traversal algorithms. Suppose that the overhead per 
forward step is bounded by some constant $c$. Then, we partition the list into blocks, 
each consisting of a sublist of $c/\epsilon$ nodes. We now apply our algorithm to a new 
list whose nodes are the blocks. The overhead per forward step is now at most $c$ per 
block, and hence at most $c/(c/\epsilon)=\epsilon$ per node. The penalty in executing 
the algorithm for blocks rather than nodes is that each forward traversal in the new 
list now translates to $c/\epsilon$ list steps, so the overhead per back-step is now 
multiplied by $c/\epsilon$. 
\end{Proof}
}

\Subsection{Full algorithm implementation using the Recycling Bin data structure}{RB-ds}
The allocation of pebbles is handled by an auxiliary data structure, denoted as the 
{\it recycling bin data structure}, or {\it RB}. The RB data structure supports the following 
operations:\\
{\it Put pebble:} put a released pebble in the RB for future use; this occurs in the simple 
back-step, in which the current position is a left child, and therefore its predecessor 
is its parent. (Back-step Case 1.)\\
{\it Get pebble:} get a pebble from the RB; this occurs in a simple forward step,
in which the successor of the node of the current position is its left child. 
(Forward-step Case 1.)\\
{\it Put list:} put a released list of pebbles - given by a pair of pointers to its head 
and to its tail -- in the RB for future use; this occurs in the non-simple forward step, 
in which the pebbles placed on a full right path should be released. 
(Forward-step Case 2.)\\
{\it Get list:} get the most recent list of pebbles that was put in the RB and is still 
there (i.e., it was not yet requested by a get list operation); this occurs in 
a non-simple back-step, in which the current position is a right child, and therefore 
its predecessor is a rightmost leaf, and it is necessary to reconstruct the right path 
from the left sibling of the current position to its rightmost leaf. It is easy to verify
that the list that is to be reconstructed is indeed the last list to be released and put 
in the RB. (Back-step Cases 2 or 3.)\\

The RB data structure consists of a bag of pebbles, and a set of lists consisting of 
pebbles and organized in a double-ended queue of lists. The bag can be implemented as, 
e.g., a stack. For each list, we keep a pointer to its header and a pointer to its tail, 
and the pairs of pointers are kept in doubly linked list, sorted by the order in which 
they were inserted to RB. Initially, the bag includes $2\log n$ pebbles and the lists 
queue is empty. 
Based on Theorem~\ref{Theorem:2logn-pebbles}, 
 the $2\log n$ pebbles will suffice for all operations.

In situations in which we have a {\tt get pebble} operation and an empty bag of pebbles, we 
take pebbles from one of the lists. For each list $\ell$ we keep a counter $M_{\ell}$ for 
the number of pebbles removed from the list. 

The operations are implemented as follows:\\
{\it Put pebble:} Adding a pebble to the bag of pebbles (e.g., stack) is trivial; it takes 
$O(1)$ time.\\
{\it Put list:} a new list is added to the tail of the queue of lists in RB, to become the
last list in the queue, and $M_{\ell}$ is set to 0. This takes $O(1)$ time. \\
{\it Get pebble:} If the bag of pebbles includes at least one pebble, return a pebble from
the bag and remove it from there. If the bag is empty, then return and remove the last 
pebble from the list $\ell$, which is the oldest among those having the minimum $M$, 
and increment its counter $M_{\ell}$. This requires a priority queue according to the 
pairs $\langle M_{\ell},R_{\ell} \rangle$ in lexicographic order, where $R_{\ell}$ is the rank of 
list $\ell$ in RB according to when it was put in it. We show below that such PQ can 
be supported in $O(1)$ time.\\ 
{\it Get list:} return the last list in the queue and remove it from RB. If pebbles were
removed from this list (i.e., $M_\ell > 0$), then it should be reconstructed in 
$O(2^{M_\ell})$ time prior to returning it, as follows. Starting with the node $v$ of 
the last pebble currently in the list, take $2^{M_\ell}$ forward steps, and whenever 
reaching a node on the right path starting at node $v$ place there a pebble obtained 
from RB using the get pebble operation. Note that this is Back-step Case 3, and according
to the analysis and claim the amortized cost per back-step is $O(\log n)$ time.

\begin{Claim}{PQ}
The priority queue can be implemented to support delmin operation in $O(1)$ 
time per retrieval. 
\end{Claim}

\RemoveICALP{
\begin{Proof}
We rely on the fact that the sequence of counters $M_\ell$ in order 
defined by the queue of lists is of a non-increasing sequence. 
Indeed, initially all $M_{\ell}$ are equal (all 0). 
Inductively, the monotonicity is preserved since when a pebble is removed from a list, 
the $M_\ell$ that is incremented is for a list $\ell$ that is the farthest list 
 having this counter. 
When a list is added, it has a counter 
of 0, and it is at the beginning of the queue. A removal of the list clearly cannot 
affect the monotonicity. Based on this property, the PQ is implemented as follows. 
All lists with the same counter $M$ are kept in a sublist (sorted by their rank), 
for each value of the counter $M$ we keep a pointer to the farthest list, and these 
pointers are linked in increasing order of respective $M$. Recall that the lists are 
organized in a queue of a doubly linked list. To support delmin operation, we take a 
pebble from the list according to the first pointer, move this list to the next sub-list,
and move this pointer to one list closer. 
\end{Proof}
}

We can conclude:
\begin{Theorem}{recycling-bin-ds}
The list pebbling algorithm using the recycling bin data structure supports $O(\log n)$ amortize 
time per back-step, $O(1)$ time per forward step, using $O(\log n)$ pebbles.
\end{Theorem}

\begin{Proof}
The theorem follows from the fact that the order in which pebbles are removed and put in 
the lists, as implemented in the recycling bin data structure, is the same as in the
list pebbling algorithm described in Section 3.2. 
\end{Proof}

\Remove{
\Section{The advanced list pebbling algorithm}{Advanced-list-pebbling}
The advanced list pebbling algorithm presented in this section supports back-steps 
in $O(\log n)$ time per step in the worst case. 
Ensuring $O(\log n)$ list steps per back-step in the worst case is obtained by processing 
the rebuilt of the missing green paths along 
the sequence of back traversal, using a new set of red pebbles. For each green path, 
there is one red pebble whose function is to progressively move forward from the 
deepest pebbled node in the path, to reach the next node to be pebbled. By synchronizing 
the progression of the red pebbles with the back-steps, we can guarantee that green 
paths will be appropriately pebbled whenever needed.

The number of pebbles used by the algorithm is bounded by $\log n$ (rather than $O(\log n)$).
This is obtained by a careful implementation and a sequence of refinements, described in the appendix.

\begin{Theorem}{RAM}
The list pebbling algorithm can be implemented on a RAM to support $O(\log i)$ time in the 
worst-case per back-step, $1+\epsilon$ time per forward step, using $\log n$ pebbles,
where $i$ is the distance from the current position to the farthest point traversed so 
far. The memory used is at most $1.5(\log n)$ words of $\log n + O(\log \log n)$ bits 
each.
\end{Theorem}

Details are given in the appendix.
}


\Remove{
\bibliographystyle{abbrv}
\bibliography{paper}

\appendix
}

\Section{The advanced list pebbling algorithm}{Advanced-list-pebbling}
The advanced list pebbling algorithm presented in this section supports back-steps 
in $O(\log n)$ time per step in the worst case. 
We present in Section~\ref{advanced-pebbling-technique} the refined pebbling 
technique, and we
describe in Section~\ref{Section:encoding} the traversal synopsis implementation.

\subsection{The advanced pebbling technique}\label{advanced-pebbling-technique}
Ensuring $O(\log n)$ list steps per back-step in the worst case is obtained by processing 
the rebuilt of the missing green paths along 
the sequence of back traversal, using a new set of red pebbles. For each green path, 
there is one red pebble whose function is to progressively move forward from the 
deepest pebbled node in the path, to reach the next node to be pebbled. By synchronizing 
the progression of the red pebbles with the back-steps, we can guarantee that green 
paths will be appropriately pebbled whenever needed.

We manage to reduce the total number of pebbles to $\log n$. Blue pebbles are saved by 
relying on recursive application of the list pebbling algorithm, and green pebbles are saved 
by delaying their creation without affecting the back-step time. The algorithm that 
realizes the following theorem is described in Section~\ref{4.1}.

\begin{Theorem}{logi-algorithm}
The list pebbling algorithm can be implemented to support $O(\log i)$ list-steps in the 
worst-case per back-step, one list-step per forward step, using $\log n$ pebbles, 
where $i$ is the distance from the current position to the farthest point traversed 
so far.
\end{Theorem}


\subsubsection{Back traversal in at most $\log n$ steps per back-step, using $2\log n$ pebbles}
\label{4.1}
As in the basic list pebbling algorithm we will use the blue pebbles placed on the path from 
the root to current position $i$, as well as 
an additional set of at most $\log n$ green pebbles that will be placed in advance and 
be converted to blue pebbles at the appropriate time. 
Additionally, we will use a new set of at most $\log n$ {\it red pebbles} that will 
serve to get to the positions of the green pebbles.

For illustration purpose, the dynamics of pebbles looks as follows: At all time, there 
are up to $\log n$ red pebbles that each advance one step per back-traversal; 
their exact number equals the number of right sub-trees that include the current position. 
Green pebbles are created when red pebbles reach certain locations, at some point each 
green pebble becomes a blue pebble, and blue pebbles are dismissed. 

When a node $i$ is a right child, then there should be a green path from $i$'th left sibling
to node $i-1$ (which is the rightmost leaf in the tree rooted at that sibling). We start 
placing the green pebbles on this path well in advance, according the following 
strategy. Let $T$ be the sub-tree of node $i$, and let $T'$ be the sub-tree of its left
sibling. As we enter via a back step into $T$ (that is, moving into the rightmost leaf 
of $T$), and start back traversing within $T$, we also start forward traversing, using 
a red pebble, the sub-tree $T'$ at the same rate. 
Whenever encountering a node in $T'$ that is on the path from the root of $T'$ to $i-1$, 
we can place there a green pebble. When the back traversal reaches node $i$, the forward 
traversal in $T'$ ends and we have all green pebbles in place. 
In the next back-step, moving from node $i$ to node $i-1$, the green pebbles in $T'$ 
will be transformed into blue pebbles.

Since every node is in at most $\log n$ sub-trees, we have at most $\log n$ such processes 
occurring in parallel, and can be implemented in a dove-tailing fashion in $\log n$ steps. 
Thus, the process consists of using up to $\log n$ red pebbles, 
and traversing each of them one step forward for every back-step, resulting with an 
overhead of at most $\log n$ steps per back-step. 

\Remove{
\begin{COMMENT}{ICALP}
\end{COMMENT}
The details of how we bound the number of pebbles to $2\log n$ is given in the
full paper. We only highlight here the considerations, by first bounding their
number to $3\log n$ and then to $2\log n$.
 
Since for each blue pebble there can be at most one green path associated with it (the
one starting at its left sibling), and since each green path is of length at most $\log n$,
this approach requires at most $\log^2 n$ green pebbles. 
Next we show that the above strategy can be implemented using at most $\log n$ green pebbles. 
We inspect more carefully the nodes on the tree that obtain green pebbles at every given 
step, and show that the total number of green pebbles that are placed preemptively (in the 
various sub-trees) is at most $\log n$.

To implement the algorithm using only $2\log n$ pebbles, we will not keep blue pebbles 
on left-subpaths in $p$, except for their first nodes. It is easy to see that every 
node on a left subpath of length $k$ starting at a node $v'$ can be reached in at 
most $k$ steps, by accessing $v'$ and moving forward. Additionally, for every right 
subpath starting a node $v$ (which is a left child of $p$) we will not keep the 
green pebble on $v$, since it can be reached in one step from $p$. It is easy to 
verify that the total number of blue and green pebbles is at most $\log n$.

The last modification can add at most $\log n$ steps to each back-step, assuring 
an implementation that uses $2\log n$ pebbles and at most $2\log n$ steps for 
every back-step. In fact, a closer analysis shows that the number of steps per 
back-step is at most $\log n$ in the worst case. First, observe that the number 
of red pebbles is the number of right sub-trees to which the current position belongs 
(this is also the number of blue pebbles). For a node whose left subpath is of 
length $k$, the number of red pebbles could be at most ($\log n - k$). Hence, the total
number of steps for moving the red pebbles, as well as reaching a node is at most 
$\log n$.

Details of the above are given in the full paper, where we also show how back-queries 
can be supported in constant time.
}

\subsubsection*{Bounding the number of pebbles to $3 \log n$}
 
\RemoveICALP{
Since for each blue pebble there can be at most one green path associated with it (the
one starting at its left sibling), and since each green path is of length at most $\log n$,
this approach requires at most $\log^2 n$ green pebbles. 
Next we show that the above strategy can be implemented using at most $\log n$ green pebbles. 
We will inspect more carefully the nodes on the tree that obtain green pebbles at every given 
step, and show that the total number of green pebbles that are placed preemptively (in the 
various sub-trees) is at most $\log n$.

\begin{Lemma}{green-pebbles-logn}
Using the above implementation, at every point in the execution the total number of green 
pebbles in the tree is at most $\log n$.
\end{Lemma}
\begin{Proof}
Recall that we have defined a {\it left-subpath} ({\it right-subpath}) as a path consisting 
of nodes that are all left children (right children). Further, considering the (blue-pebbled) 
path $p$ from the root to node $i$, we say that $v$ is a left-child of $p$ if it has a 
right sibling that is in $p$ (that is, $v$ is not in $p$, it is a left child, and its parent 
is in $p$ but not the node $i$). 

Green pebbles are placed preemptively on right-subpaths that begin at left children of $p$. 
We claim that the number of green pebbles placed preemptively 
on the right-subpath starting at a node $v$, is at most the length of the mirror sub-path plus 1.
(Recall that the mirror sub-path is the left-subpath in $p$, starting at the right sibling 
$v'$ of $v$). The lemma will follow immediately since the sum of lengths of $L$ left subpaths 
along $p$ is at most $\log n - L$.

Consider some left-subpath of length $k$ starting at a node $v'$, and ending at a node 
$u$. We show inductively (on the length $k$) that at the right-subpath starting at $v$,
the left sibling of $v'$, has at most $k+1$ pebbles. The base, $k=0$ follows from the 
fact that the traversal did not yet enter into the sub-tree of $v'$, and hence the only
green pebble in the sub-tree of $v$ is on $v$ itself.  To prove the induction step, 
consider the case that the left-subpath is augmented by a new node $u+1$ 
(left child of node $u$). The number of back-steps taken since adding $u$ to the 
path till this point is at most $S(u+1)$, the size of the sub-tree whose root is $u+1$,
since the traversal was taken over the right sub-tree of $u$. The traversal 
necessary from placing the $k$'th pebble to placing the $k+1^{st}$ pebble on the 
right-subpath starting at $v$ can be verified to be $S(u+1)$.  The claim follows 
from the induction hypothesis.
\end{Proof}
}

The lemma implies that at most $3\log n$ pebbles are sufficient. 

\subsubsection*{Using at most $2 \log n$ pebbles}

To implement the algorithm using only $2\log n$ pebbles, we will not keep blue pebbles 
on left-subpaths in $p$, except for their first nodes. It is easy to see that every 
node on a left subpath of length $k$ starting at a node $v'$ can be reached in at 
most $k$ steps, by accessing $v'$ and moving forward. Additionally, for every right 
subpath starting a node $v$ (which is a left child of $p$) we will not keep the 
green pebble on $v$, since it can be reached in one step from $p$. It is easy to 
verify that the total number of blue and green pebbles is at most $\log n$.

The last modification can add at most $\log n$ steps to each back-step, assuring 
an implementation that uses $2\log n$ pebbles and at most $2\log n$ steps for 
every back-step. In fact, a closer analysis shows that the number of steps per 
back-step is at most $\log n$ in the worst case. First, observe that the number 
of red pebbles is the number of right sub-trees to which the current position belongs 
(this is also the number of blue pebbles). For a node whose left subpath is of 
length $k$, the number of red pebbles could be at most ($\log n - k$). Hence, the total
number of steps for moving the red pebbles, as well as reaching a node is at most 
$\log n$.

(Note: there is an additional delay in the beginning of a traversal of a red pebble.
To reach a node $v$ that is a left child of $p$, it first requires starting at the 
beginning of the left sub-path ending at the right sibling of $v$. 
This delay can be overcome during the traversal of the red pebble.) 

A drawback of this modification is that back-query is not supported in constant time. 
Further, the lack of pebbles within a blue left sub-path would imply more expensive 
implementation of shrinking the data structure, discussed below.

We show next that this problem can be overcome by maintaining recursively a data 
structure over the blue left sub-path, as follows.

\subsubsection*{Supporting back-query in constant time  }

This is done by considering each left-subpath as a linked list, and implementing 
backward steps along a left-subpaths by using the above algorithm recursively.
For such implementation, a left-subpath of length $k$ requires $\log k$ pebbles.
We show that such pebbles are available without increasing the total number of pebbles.

\begin{Claim}{left-subpath}
Consider some left-subpath of length $k$ starting at a node $v'$, and ending at a 
node $u$. The number of green pebbles placed preemptively on the right-subpath 
starting at $v$, the left sibling of $v'$, is at most $k-\log k$.
\end{Claim}

\begin{Proof}
The left-subpath starting at $v'$ ends at node $u=(v'+k)$. The back traversal executed 
while the left-subpath in $p$ starting at $v'$ grows from $0$ to length $k$ begins at 
the rightmost leaf in the sub-tree of $v'$, and ends at the right sibling of $u$.
The length of this back-traversal is hence $S(v')-S(u)-k$.
While conducting this back traversal, the green pebbles were placed along a forward 
traversal starting at the parent of $v'$. The number of nodes in the right-subpath 
starting at $v$, the left sibling of $v'$, is indeed $k-\log k$.
\end{Proof}

\subsubsection{Using at most $\log n$ pebbles}
We impose a delay in the creation of the red pebbles. Specifically, red pebbles will 
be created at a node only when reaching the sub-tree of the left child of its right 
sibling (if such right sibling exists).
The red pebble at the root of $T'$ will be created when reaching the sub-tree of 
the left child of the root of $T$.

Observe that (1) at every step, the number of red pebbles is the number of left-paths;
(2) for each right-path of green pebbles, the number of green pebbles will be reduced 
by one. Since the number of green right-paths equals the number of left-paths, the 
total number of pebbles is at most $\log n$.

We still need to guarantee that when reaching the root of $T$, the path from the root 
of $T'$ to its right leaf is already pebbled (by green pebbles). To compensate for 
the delay in creating the red pebble, once created the red pebbles are to be moved 
in double pace. That is, for every back-step each of the red pebbles is to be moved 
two forward steps. To see that the time per step is still bounded by $\log n$, note 
that the number of red pebbles is bounded by $\log n \over{2}$, the maximum 
possible number of left paths.

\subsubsection{Forward traversal in $O(1)$ time and no additional list-steps}

When implementing forward traversal, blue pebbles will be updated to always be in 
 their designated positions. 
Green and red pebbles will be eliminated as necessary, 
 so as to maintain that the blue pebbles use a total of $\log n$ pebbles. 
Thus, no additional list-steps are needed in order to maintain the pebbles in their
 appropriate positions, during forward steps. 
The change in pebble positions is obtained by recording new positions as they are 
 visited during the forward traversal.
The only overhead is incurred by maintaining the data structures necessary to identify
 the appropriate pebble locations, namely the recycling-bin data structure and the virtual
pre-order tree data structure.

The implementation of the forward traversal is essentially rolling back the back 
traversal described above, with the following modifications. A red pebble that is to 
be moved backward is only moved virtually and will in fact remain in place 
(a counter can be added to explicitly represent the virtual location). The rolling 
back of a step in which a green pebble is created is implemented by removing the 
green pebble and moving the red pebble that created it into its position. 
Otherwise, rolling back a step in which a red pebble is created is removing the red 
pebble; rolling back a step in which a blue pebble is eliminated is by creating the 
blue pebble; rolling back a step in which a green pebble becomes blue is by having 
the blue pebble becoming green. 

Since for every step only one new blue pebble can be created, we can eliminate only 
 one pebble (green or red) per step. 
Thus, occasionally we will take a lazy approach in rolling back and deferring some 
 elimination to other steps.  
This results with the cost of two operations per forward step (creation of a blue pebble 
 and one elimination). 
Additional cost per step is the change of blue pebbles into green pebbles. 
This is implemented in a lazy fashion as well - one change per step. 

The correctness of this implementation follows from the previous discussion. 
The total cost per forward step is (a small) constant in the worst case, and 
 can be also bounded by $1+\epsilon$ amortized cost per forward step, as will 
 be shown  below.

\RemoveICALP{

\subsubsection{Back traversal in $O(\log i)$ for the $i$'th back-step }
It is desirable that when executing a small number of back-steps, their cost 
would be constant. More generally, we would like the cost of a back-step to be a 
function of the number of back-steps executed. That is, that the $i$'th back-step 
costs $O(\log i)$, rather than $O(\log n)$. When starting back-steps after a 
sequence of $n$ forward steps, where $n$ is a power of $2$, using the above algorithm, 
then in fact we do already obtain a cost of at most $\log i$ per back-step. Indeed, the 
cost is proportional to the number of red-pebbles and this number is at most $\log i$ 
for the $i$'th back-step, since the largest tree affected when traversing $i$ 
steps backward is of size at most $2i$. However, using the above algorithm for 
arbitrary $n$, starting back traversal after moving forward $n$ steps could cost 
$\Theta(\log n)$ per back-step. 

We modify the implementation of back-step so as to obtain cost of $O(\log i)$ 
regardless of the point in which the back traversal begins, by creating red pebbles 
mostly in a tree whose size is at most $4i$. Creation of red pebbles in most 
of the higher nodes is delayed to future back-steps. 

\RemoveICALP{
\begin{Claim}{smallest-sub-tree}
Consider the smallest sub-tree $T$ that includes both nodes representing positions $n$
and $n-i$, and let $0 < j\le i$. (Note that this tree could be of size up to $2n$ even
for very small $i$). Then, the number of red-pebbles required for the node representing
$n-j$ within the tree is at most $\log i \over 2$$+1$.  
\end{Claim}

\begin{Proof}
Since the tree $T$ is the smallest that includes the sequence of nodes representing 
positions $n-j$, $0<j\le i$, then the right child of the root of this tree represents
a node $n-j'$ in the sequence, for some $0<j' \le i$. We consider separately the two 
parts of the sequence.

The suffix of the sequence, $n-j'-1,\ldots ,n-i$, begins at the right leaf of the 
left sub-tree of $T$ (the situation is similar to back-traversal from a position $n$ 
that is power of $2$), and is fully contained in a sub-tree $T'$ of size at most $2i$.
Within $T'$, the number of red pebbles is at most $\log i + 1 \over 2$, and no 
additional red pebble is required since the root of $T'$ is on a right path from the 
left child of the root of $T$. 

The prefix of the sequence, $n,\ldots,n-j'$, consists of two sub-sequences, one on a 
left path $p$ starting at the right child of the root of $T$, and the other fully 
contained in a sub-tree $T''$ whose root is on $p$, and whose size is at most $2i$.
One pebble is required for $p$, and within $T''$, the number of red pebbles is at 
most $\log i + 1 \over 2$, totaling at most $\log i +1 \over 2$$ + 1$.
\end{Proof}

The algorithm is as follows. Let $R$ be the smallest right sub-tree that fully 
contains the tree $T$ (as a sub-tree) from the claim. Then, at every back-step only 
red pebbles within $R$ are created or moved. Each pebble except for the highest 
pebble is moved two steps per every back-step, as before. The highest pebble is 
moved at up to double pace, that is up to 4 steps for every back-step. 
The double pace covers the delay this pebble has had, when it was not in the tree $R$ 
(when back traversing within the right sub-tree of $R$). Creation or move of 
red-pebbles outside $R$ is delayed until a later stage. 

>From the claim it follows that the number of red-pebbles touched at every back-step 
is at most $\log i + 1 \over 2$$ + 2$. Each pebble, except for the highest pebble,
is moved two steps for every back-step; the highest pebble is moved at most 4 steps 
for every back-step. The total number of steps per back-step is therefore at 
most $\log i + 7$.

It remains to show that the delay of high pebbles is properly compensated by the 
double pace approach. It suffices to show that the lowest delayed red pebble is in 
place when first needed. Let $L$ be the sub-tree whose root is the left sibling of 
the root of $R$. The size of $L$ is at least $i$, and once entering $L$, the lowest 
delayed red pebble starts moving in double pace. Therefore, when finishing the traversal 
through $L$, the green pebble created by that delayed pebble would already be in place. 
}
}

\Remove{
\begin{COMMENT}{ICALP}
\end{COMMENT}
\section{Back traversal in $O(\log i)$ for the $i$'th back-step }
It is desirable that when executing a small number of back-steps, their cost 
would be constant. More generally, we would like the cost of a back-step to be a 
function of the number of back-steps executed. That is, that the $i$'th back-step 
costs $O(\log i)$, Rather than $O(\log n)$. When starting back-steps after a 
sequence of $n$ forward steps, where $n$ is a power of $2$, using the above algorithm, 
then in fact we do already obtain a cost of at most $\log i$ per back-step. Indeed, the 
cost is proportional to the number of red-pebbles and this number is at most $\log i$ 
for the $i$'th back-step, since the largest tree affected when traversing $i$ 
steps backward is of size at most $2i$. However, using the above algorithm for 
arbitrary $n$, starting back traversal after moving forward $n$ steps could cost 
$\Theta(\log n)$ per back-step. 

We modify the implementation of back-step so as to obtain cost of $O(\log i)$ 
regardless of the point in which the back traversal begins, by creating red pebbles 
mostly in a tree whose size is at most $4i$. Creation of red pebbles in most 
of the higher nodes is delayed to future back-steps. 

\begin{COMMENT}{ICALP}
\end{COMMENT}
Details are given in the full paper.
}

\subsubsection{Super-nodes for $\epsilon$ overhead in forward steps}

\begin{figure}[htb]
\centerline{\psfig{figure=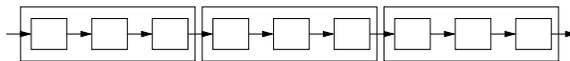,height=0.3in,width=3.0in}}
\caption{Obtaining $\epsilon$ overhead by using super-nodes}
\label{fig:supernode}
\end{figure}

The overhead in forward step was shown to be $O(1)$ in the worst case.
We show how to reduce the amortized time overhead per forward step to 
 $\epsilon$, for arbitrary small $\epsilon > 0$.
Let $c>0$ be a constant so that each forward step takes at most $c$ time.
Given a list $L$ of length $n$, we define a virtual list $L'$ over $L$, as follows
 (see Figure~\ref{fig:supernode}).
Each node in $L'$ is a super-node, representing a group of $c/\epsilon''$ consecutive
 nodes in $L$, where $\epsilon'' < \epsilon$ will be defined below.
We keep a list traversal synopsis over $L'$, so that whenever a pebble is allocated
 to a super-node in $L'$, we will instead allocate the pebble to the first node
 of $L$ within that super-node.

A forward step in $L'$ from a super-node $v'$ to a subsequent 
 super-node $u'$ occurs only when there is a forward step in $L$ from the last 
 node in $v'$ to the first node in $u'$. 
Since the time per forward step in $L'$ is at most $c$, the amortized time per 
 forward step in $L$ is at most $\epsilon' + c/ (c/\epsilon'') = \epsilon' + \epsilon'' = \epsilon$
 where $\epsilon'$ is the overhead per step due to managing the super-nodes, and 
 $\epsilon''$ defined as $\epsilon - \epsilon'$. 

A backward step in $L$ can be of two types. If the back-step is between two nodes
 belonging to the same super-node, then it is executed by moving forward at most
 $c/\epsilon'' = (1/\epsilon)$ steps from the beginning of the shared super-node; 
 no back-step occurs in $L'$ in this case.
If the back-step is between two nodes belonging to a different super-nodes, then a
 back-step in $L'$ is executed, and the required node is reached by 
 moving forward $O(1/\epsilon)$ steps from the beginning of its super-node. 
Executing a back-step in $L'$ involves a sequence of forward list-steps  in $L'$; each
 such list-step is executed by having $c/\epsilon''$ forward steps in $L$'.
Hence, the time per back-step is now multiplied by a factor of $O(1/\epsilon)$.

\Subsection{Traversal Synopsis implementation}{encoding}
A full implementation of the advanced list pebbling algorithm requires identification of the 
 positions of red pebbles, with the constraint that the tree on which they move is 
 only virtual.

\begin{Theorem}{RAM}
The list pebbling algorithm can be implemented on a RAM to support $O(\log i)$ time 
 in the worst-case per back-step, where $i$ is the distance from the current position 
 to the farthest point traversed so far. 
Forward steps are supported in $O(1)$ time in the worst case, $1+\epsilon$ amortized time 
 per forward step, and no additional list-steps, using $\log n$ pebbles.
The memory used is at most $1.5(\log n)$ words of $\log n + O(\log \log n)$ bits 
each.
\end{Theorem}

\RemoveICALP{
\begin{Proof}
A straightforward implementation of the advanced pebbling algorithm described in
 Section~\ref{advanced-pebbling-technique} requires a full tree 
representation. As we aim to small space data structure, this tree will only be 
virtual and we'll show how to implement with logarithmic size data structures.


The data structure, dubbed the pebble tree, consists of a binary tree, which is 
induced by the sub-tree of the full binary tree described above. At every point of time, 
the pebble tree consists of nodes positioned with blue or green pebbles, and additional 
leaves representing the red pebbles. These leaves are the right children of the nodes 
with the corresponding green pebbles.

Each node in the tree has three pointers: to its left child, to its right child, and to 
its parent. Each node in the tree had a record with the following information: 
the node's position in the tree ($\log n$ bits), pointer id ($\log \log n$ bits), and delay 
for nodes representing red pebbles ($\log n$ bits), which is used in the forward step. 

Additionally, we keep an array, which keeps pointers to the nodes representing 
the red pebbles in the pebble tree. Finally, we keep two pointers, to the nodes 
representing the current position (the node representing the last blue pebble), and to 
the node representing the last green pebble of a full green path (that is, a node which 
is a leaf in the full tree).

The pebble tree consists of at most $\log n$ nodes.  
%
%
%
%
The number of red nodes is at most $\log n \over 2$, and hence this is the size of 
the array. Representing the pebble tree structure and all pointers involved 
requires $O(\log \log n)$ bits per pointer. Each node requires $\log n$ bits per position, 
and each red node requires additional $\log n$ bits for the delay field. 

In total we have at most $\log n$ words of size $\log n + O(\log \log n)$, of which at 
most $1\over 2$$\log n$ words require additional words of size $\log n$, 
and an array consisting of $\log n$ words of size $O(\log \log n)$. 

Resulting with total of $1.5 \log n (\log n + O(\log \log n))$ bits.
\end{Proof}
}

\Remove{
 \subsection{Extension to arbitrary $k$}
The virtual tree data structure can be extended to support the full memory-time trade-off.
The same virtual tree serves both the case of having $O(k)$ pebbles with $O(k n^{1/k})$ 
time, and the case of having $O(k n^{1/k})$ pebbles and $O(k)$ time.  
It  consists of a virtual pre-order tree data structure of depth $k$ and degree $n^{1/k}$.  
As for the binary tree, in both cases we maintain a blue pebble path from the root to the 
current position (of length at most $k$). The difference is in the placement of the green
pebbles.


For the case of $O(k)$ pebbles and $O(k n^{1/k})$ time per back-step, we have the green
paths begin only at the nearest left sibling of each node with a blue pebble. As in the
binary case, they consist of right sub-paths and their lengths is according to the mirroring
property.
As in the binary case, the total number of green pebbles is at most the total number of
blue pebbles, which is at most $k$. Hence, we have at most $2k$ pebbles in the tree.
The time per back-step is dominated by the time it takes to reconstruct a new green path.
Consider a back step from a root $v$ of a sub-tree $T$. 
Constructing the green path that starts at the parent of $v$ involves forward traversal 
through at most $n^{1/k}$ sub-trees identical to $T$. As in the binary case, the
construction time is amortized against the sequence of back-steps from the right-most
leaf of $T$ till the root of $T$. 
Since each node is in at most $k$ different trees, and hence each back-step is amortized 
against at most $k$ such green path constructions, the amortized time per back step is
$O(k n^{1/k})$. 
Red pebbles can be used as in the binary case to enable worst-case performance, 
increasing the total number of pebbles to at most $3k$. Similar techniques to the
binary case can be used to reduce the number of pebbles to at most $k$.
Forward steps take $O(1)$ each as before. 

\begin{figure}[htb]
\centerline{\psfig{figure=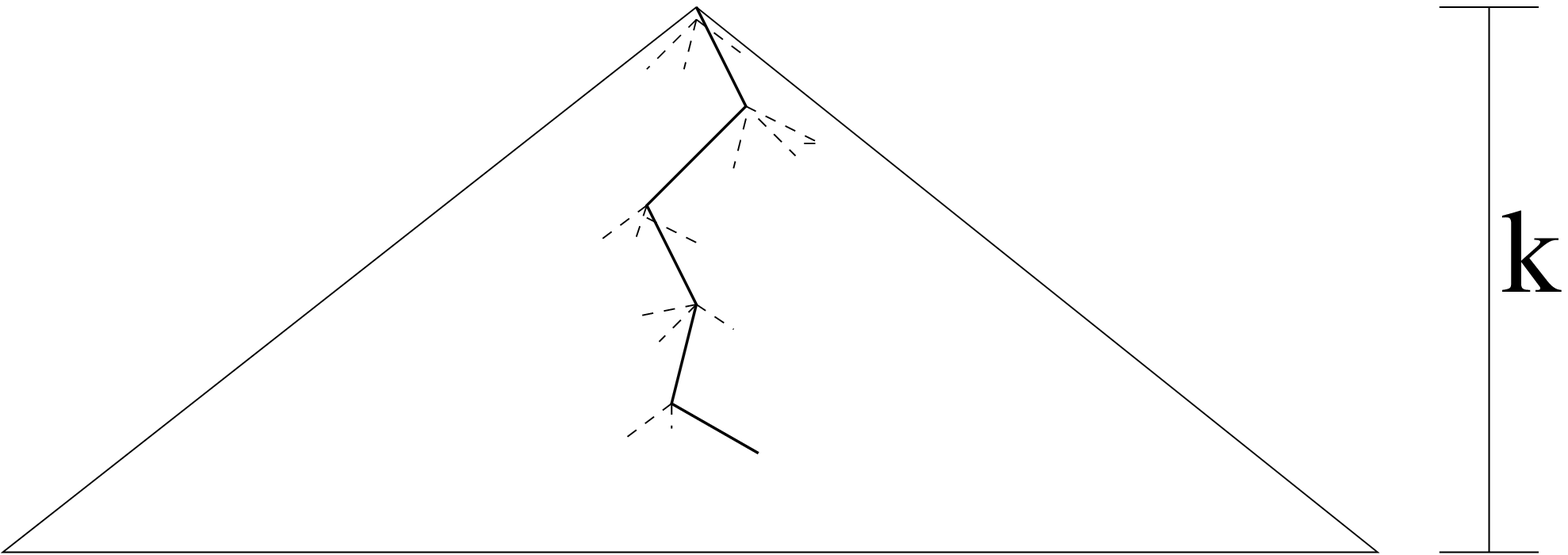,height=1.0in,width=2.0in}}
\caption{The generalized tree}
\label{fig:tradeoff}
\end{figure}

\begin{COMMENT}
Place figure
\end{COMMENT}

For the full trade-off in which the number of pebbles is $O(k n^{1/k})$ and the time per
back-step is $O(k)$, we use the same virtual tree of depth $k$. We place blue and green 
pebbles as before, but in addition we place green pebbles at all left siblings of each 
node with a blue or green pebble. 
The number of pebbles is at most $n^{1/k}$ times the number of pebbles in the previous
case, that is at most $2k n^{1/k}$.
The amortized time per back-step here is $O(k)$. Indeed, the same amortization argument
of the previous case applies here, except that constructions of green paths only involve
forward traversal of one sub-tree. 
Worst case performance is again obtained using red pebbles, and $O(1)$ time per 
forward step as well as reducing the number of pebbles is obtained as before.
}

\RemoveICALP{
\subsection{Growing the virtual tree for an unbounded sequence}

Suppose that we have a list traversal synopsis on a virtual pre-order
 tree $T$ of $n$ nodes.
When moving forward on an unbounded list from node $n$ to node $n+1$, the
 tree $T$ can no longer be used for the list traversal synopsis.
Instead, a (virtual) tree $T'$ of size $2n+1$ is to be used.
Note that the blue path of node $n$ in $T$ consists of the right path in $T$.
Our objective is to obtain the blue path of $n+1$ in $T'$.


We note that $T$ is identical to the left sub-tree 
 $T''$ of $T'$, and it is easy to see that the pre-order numbering of each 
 node in $T''$ is one more than the pre-order numbering of the corresponding 
 node in $T$.
Thus, node $n+1$ is the rightmost leaf of $T''$, and the blue path of $n+1$
 in $T'$ consists of the root of $T'$ along with the right path of $T''$.
By moving each blue pebble to its next position, it is brought in $T''$ to
 the position that corresponds to its position in $T$, hence the entire
 blue path of $n+1$ is obtained, except for the root of $T$.

Thus, the blue path for node $n+1$ in $T'$ can be obtained by moving each 
 blue pebble to its next position, and adding a new blue pebble
 to the root of $T'$. 
The amortized cost of computing the new blue path is $\log n / n$, and 
 such update can be prepared in advance in a straightforward manner, so as
 to keep the time per forward step $O(1)$ in the worst case.

\paragraph{Reducing the number of pebbles during back traversal}
The number of pebbles can also be kept at all time $O(\log n)$, where $n$ is the
 current position, even if at some point in the past the virtual tree grew to
 arbitrary size $N \gg n$.
Indeed, the blue path consists of a left path of some length $n_1$, and some
 blue path within a subtree $S$ of length at most $2n_2$, so that $n_1+n_2 = n$.
As in the advanced algorithm using $\log n$ pebbles, the left blue path can
 be substituted with a recursive instance of a list traversal synopsis of size
 $O(\log n_1)$.
Since the list traversal synopsis of the subtree $S$ is of size $O(\log n_2)$,
 we have a total of $O(\log n)$ pebbles. 

}

\section{Obtaining a full time-memory trade-off}
\begin{figure}[htb]
\centerline{\psfig{figure=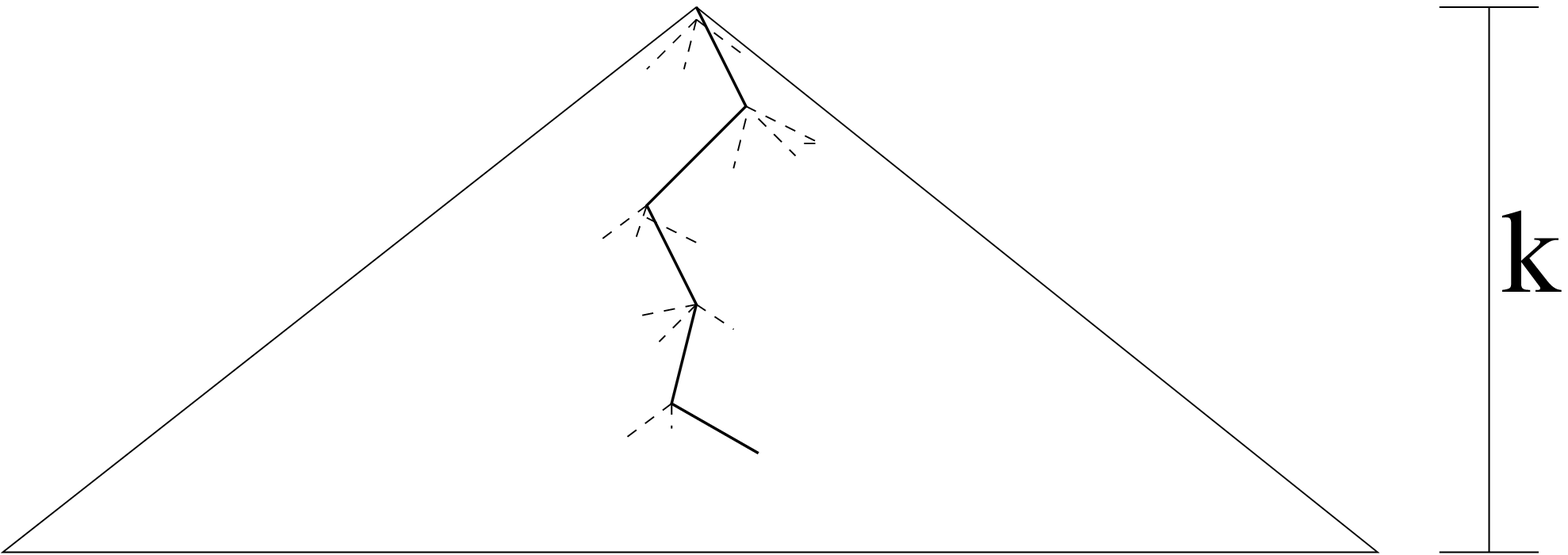,height=1.0in,width=2.0in}}
\caption{The $k n^{1/k}$-ary virtual pre-order tree, supporting $k$ vs.\ $kn^{1/k}$ trade-off.}
\label{fig:k-ary}
\end{figure}

In order to support a full memory-time trade-off, we extend the virtual 
 pre-order tree data structure from a binary tree into a 
 $k n^{1/k}$-ary tree
 (see Figure~\ref{fig:k-ary}).
The same virtual $k n^{1/k}$-ary tree serves both the case of having $O(k)$ 
 pebbles with $O(k n^{1/k})$ 
time, and the case of having $O(k n^{1/k})$ pebbles and $O(k)$ time, 
but with different placements of pebbles.  
It consists of a virtual pre-order tree data structure of depth $k$ and degree $n^{1/k}$.  
As for the binary tree, in both cases we maintain a blue pebble path from the root to the 
current position (of length at most $k$). The difference is in the placement of the green
pebbles.


For the case of $O(k)$ pebbles and $O(k n^{1/k})$ time per back-step, we have the green
paths begin only at the nearest left sibling of each node with a blue pebble. As in the
binary case, they consist of right sub-paths and their lengths is according to the mirroring
property.
As in the binary case, the total number of green pebbles is at most the total number of
blue pebbles, which is at most $k$. Hence, we have at most $2k$ pebbles in the tree.
The time per back-step is dominated by the time it takes to reconstruct a new green path.
Consider a back step from a root $v$ of a sub-tree $T$. 
Constructing the green path that starts at the parent of $v$ involves forward traversal 
through at most $n^{1/k}$ sub-trees identical to $T$. As in the binary case, the
construction time is amortized against the sequence of back-steps from the right-most
leaf of $T$ till the root of $T$. 
Since each node is in at most $k$ different trees, and hence each back-step is amortized 
against at most $k$ such green path constructions, the amortized time per back step is
$O(k n^{1/k})$. 
Red pebbles can be used as in the binary case to enable worst-case performance, 
increasing the total number of pebbles to at most $3k$. Similar techniques to the
binary case can be used to reduce the number of pebbles to at most $k$.
Forward steps take $O(1)$ each as before. 

For the full trade-off in which the number of pebbles is $O(k n^{1/k})$ and the time per
back-step is $O(k)$, we use the same virtual tree of depth $k$. We place blue and green 
pebbles as before, but in addition we place green pebbles at all left siblings of each 
node with a blue or green pebble. 
The number of pebbles is at most $n^{1/k}$ times the number of pebbles in the previous
case, that is at most $2k n^{1/k}$.
The amortized time per back-step here is $O(k)$. Indeed, the same amortization argument
of the previous case applies here, except that constructions of green paths only involve
forward traversal of one sub-tree. 
Worst case performance is again obtained using red pebbles, and $O(1)$ time per 
forward step as well as reducing the number of pebbles is obtained as before.



\section{Reversal of program execution}
A unidirectional linked list can represent the execution of programs. 
Program states can be represented as nodes in a list, and a program step is represented by a
directed link between the nodes representing the appropriate program states.
Since typically program states cannot be easily reversed, the list is in general
unidirectional.

Consider a linked list the represents a particular program. 
Moving from a node in the list back to its preceding node is equivalent to 
 reversing the step represented by the link.
Executing a back traversal on the linked list is hence equivalent to rolling back
the program. 
Let the sequence of program states in a forward execution be $s_0, s_1, \ldots, s_T$. 
A {\em rollback} of a program of some state $s_j$ is changing its state to the 
preceding state $s_{j-1}$. 
A rollback step from state $s_j$ is said to be the {\em $i$'th rollback step}
if state $s_{j+i-1}$ is the farthest state that the program has reached so far.

We show how to efficiently support back traversal with negligible overhead to 
forward steps.
The following theorem follows directly from Theorem~\ref{Theorem:RAM}.

\begin{Theorem}{program-reversal}
Let $P$ be a program, using memory of size $M$ and time $T$.
Then, at the cost of recording the input used by the program, and increasing
 the memory by a factor of $O(\log T)$ to $O(M \log T)$, the program can be 
 extended to support arbitrary rollback steps as follows.
The $i$'th rollback step takes $O(\log i)$ time in the worst case, while
 forward steps take $O(1)$ time in the worst case, and $1+\epsilon$ amortized 
 time per step. 
\end{Theorem}

\subsection{Program rollback with delta encoding}\label{section:delta-encoding}
Program rollback has important applications, including debuggers and backward
 simulations. 
Traditional methods of supporting rollback is by check-pointing particular program 
 states. 
However, it is customary to also utilize the fact that changes in a program state 
 during a single program step is substantially smaller than the size of the 
 program state. 
Indeed, rather than just recording program states, one can record the differences 
 needed in program states in order to convert them into their preceding states, 
 often called {\em delta-encoding}.
This results with better memory utilization, at the cost of additional overhead
 per forward step.

Suppose that the delta-encoding is smaller than the program state by a factor of $\ell$.
Then, after $\ell$ steps the accumulated delta-encoding is about the size of 
 a single program state.
The rollback method of Theorem~\ref{Theorem:program-reversal} can be enhanced with
 the delta-encoding method, by utilizing an extension of the super-node technique 
 described above.
Each super-node consists of $\ell$ nodes in the original list. 
Additionally, the last $\ell$ steps are fully kept using delta-encoding, 
  adding at most the size of a single program state.


Forward steps are implemented as before, except that the oldest step kept in 
 delta-encoding is removed, and the delta-encoding of the new step is added.
The cost of the $i$'th back-steps is $O(\log (i/\ell))$ in the worst case. 
By delaying the maintenance of list traversal synopsis, each of the first $\ell$ 
 back-steps can be implemented in $O(1)$ time, using the available delta-encoding.
Still, for any $i > \ell$, the sequence of $i$ back-steps will take $O(i\log (i/\ell))$ 
 in the worst case. 

%
%

\RemoveICALP{
\subsection{Hash chains}

Let $h$ be a one-way cryptographic function.
A hash chain for $h$ is a sequence of hash values $v_0,v_1,\ldots, v_n$, obtained by
 repeatedly applying $h$, starting with a secret random seed $s$.
In particular, $v_0 = s$, and for all $i>0$, $v_{i} = h(v_{i-1})$.

When the seed $s$ is known, then the entire chain can be easily computed. 
However, for a party that knows $v_i$ but does not know any $v_j$ for $j<i$, the task
 of computing $v_{i-1}$ is intractable.
On the other hand, having given $v_{i-1}$, it can easily verify that $v_{i} = h(v_{i-1})$.
Hash chains are attractive in their abilities to provide a low-cost, long sequence of such 
 verification steps, where each verification involves a back step along the list representing
 the hash-chain.
Thus, the application of a hash chain is quite similar to a full program rollback, except
 that rather than keeping program states, it is sufficient to keep hash values, along with
 the hash function. 

Our list-traversal synopsis algorithm provides efficient processing of hash-chains.
By keeping $k$ hash values, it enables to get each preceding hash value 
 in $O(k n^{1/k})$ time in the worst case.
By keeping $k n^{1/k}$ hash values, the time per preceding hash value is $O(k)$
 in the worst case.
As a particular case, by keeping $\log n$ hash values, the time to obtain a preceding hash 
 value is $O(\log n)$.

The cryptographic applications of hash chain include password authentication~\cite{lamport81},
 micro-payments~\cite{Rivest:Shamir:96}, forward-secure signatures~\cite{Itkiss-Reyzin,Kozlov-Reyzin},  
 and broadcast authentication protocol~\cite{Perrig}.
Such applications and others can benefit from efficient hash-chain processing, especially in
 memory-challenged platforms such as smart-cards.
}

\Remove{
The potential benefits of using hash chains for cryptographic protocols is not new. 
It was proposed by Lamport for password authentication back in 1981 
(Password Authentication with insecure communication, CACM 24:11, 1981, 
pp 770-772), and more recently by Rivest and Shamir in 1996 for their PayWord 
micro-payment scheme (http://citeseer.nj.nec.com/rivest96payword.html). 
There are other suggestions in the literature, including the very recent ones for 
forward-secure signatures by Itkis and Reyzin 
(http://citeseer.nj.nec.com/itkis01forwardsecure.html) and by Kozlov and Reyzin 
(http://citeseer.nj.nec.com/kozlov02forwardsecure.html), and for 
broadcast authentication protocol by Perrig et al 
(http://citeseer.nj.nec.com/perrig02tesla.html). Some of the recent work are
in fact motivated by the recent work similar (and independent) to ours by 
Coppersmith and Jacobsson (http://citeseer.nj.nec.com/coppersmith02almost.html)
for efficiently handling hash chains (which does not handle more general program rollback)
}

\Section{Traversals of Trees and other graphs}{trees}

We discuss implementations for other linked structures. 
For every structure (tree, dag, graph), a (forward) traversal defines a linked list, 
 which we call {\em traversal list}, from the starting point (``root'') to the current 
 position. 
This traversal list consists of the sequence of nodes in order of traversal (possibly 
 with repetitions in case of a general graph). 
A back-step from a node in the traversal list is always defined as getting to the 
 preceding node in the traversal list.

Since the back-step is always implemented by first going to a certain pebble and 
 then moving forward from the pebble, having in-degree larger than one has no affect
 on the algorithm. 
We describe the traversal for directed trees. 
Extensions to general directed graphs is straightforward. 

Assume that we are given a directed tree, in which edges are directed from 
nodes to their children. A (forward) traversal on the tree is to move from a node 
to one of its children. A back-step from a given node is moving to its parent. 
Without additional information, a back-step can only be done by starting at the root 
and following the footsteps of the forward traversal (this can only be done under 
certain assumption to be discussed later). This requires only 2 pebbles, but requires 
time proportional to the depth of the current position.

Our objective is to support effective back-steps as well as back-queries, using small 
space. As for the case of list, a trivial support for back traversal is to add 
a back edge for every edge. A more effective approach is to add trailing back pointers 
as we go (that is, when positioned at node i we have a back path from i to the root).

Our solution for a directed list extends quite naturally to a directed tree. 
We consider two types of trees.\\
{\em Explicit}: When positioned at a node v, then given the identity of a node $u$ 
 that is a descendant of $v$, it is possible to determine in constant time which of 
 the children of $u$ is an ancestor of $v$.\\  
{\em Implicit}: Trees that are not explicit. 
When positioned at a node $v$, if $v$ has $r$ children, then 
%
%
$\log r$ bits can be used to identify each child in constant time 
(e.g., by taking the ranks in a lexicographic order of their names).

Note that the trivial solution for (expensive) back-steps using only 2 pebbles only 
 works for an explicit tree. For an implicit tree, a simple correction of the 
 algorithm is to maintain, for every node along the path from the root to the 
 current position, the identifying information regarding which child should be 
 traversed next. 
For a node $v$,  the number of bits required for this identifying information is 
 $R(v) = \sum_{u\ is\ an\ ancestor\ of\ v}{\log d(u)}$, where $d(u)$ is the number 
 of children of $u$. For a binary tree, $R(v)$ is simply the depth of $v$.

\Paragraph{Traversals of explicit trees}

As we traverse forward in the tree, we maintain a data structure corresponding to 
the list defined by the traversal. Specifically, if $p$ is the path from the root to 
the current position, then the current data structure corresponds to a traversal 
along $p$. In the implementation of a back-step or of a back query, when moving forward 
from a pebble the identity of the child to which we need to move forward is determined 
by the property of explicit trees. 

The only possible complication that one could anticipate is when having a sequence 
of back-steps, followed by moving forward along a different path. For the implementation 
that guarantees $\log n$ cost per back-step, the data structure after returning 
to a node is similar to what it would be when arriving at the node in the first 
time, except for possibly pebbles that were added during the doubling phase, and are 
now redundant.

For the implementation that guarantees $\log i$ cost for the $i$'th back-step, 
the situation is a bit subtler, since the delay while backtracking should be 
accounted for. Fortunately, the delay only affects pebbles that are quite close 
to the root, and are not affected by the backtrack sequence. 

\Paragraph{Traversals of implicit trees}

The algorithms for explicit trees works also for implicit trees, except that we need 
to resolve the identity issue as discussed above. By maintaining this information, 
using $R(v)$ additional bits when positioned at node $v$, we can indeed implement 
the algorithm described for explicit trees.

Comment: Any hierarchical structure that identifies nodes by their logical path 
is typically of explicit type. For instance: a file descriptor or a URL description. The 
``dot-dot'' operation is in fact going up a tree, and the explicit algorithm discussed 
above enables effective support of this operation even for implementations that only 
use unidirectional links.

\Section{Conclusions}{conclusions}
We presented efficient pebbling techniques, based on a novel virtual pre-order tree
 data structure, that enable compact and efficient list traversal synopses. 
These synopses support effective back-traversals on unidirectional lists, trees,
 and graphs with negligible slowdown in forward steps.
In addition to straightforward applications to arbitrary traversals on unidirectional 
 linked structures, we derive a general method for supporting efficient roll-back 
 in arbitrary programs, with small memory overhead and virtually no effect on their
 forward steps.
Other applications include memory- and time-efficient implementations of hash-chains,
 with full time-space trade-off. 



\Remove{
\begin{figure}[htb]
\centerline{\psfig{figure=sw.ps,height=1.0in,width=2.0in}}
\caption{The skeleton data structure}
\label{fig:sw}
\end{figure}

\begin{figure}[htb]
\centerline{\psfig{figure=tradeoff.ps,height=1.0in,width=2.0in}}
\caption{$\Theta(n)$ time-memory trade-off}
\label{fig:tradeoff}
\end{figure}
}


\Remove{
\appendix
 
\RemoveICALP{
\section{Back traversal in $O(\log i)$ for the $i$'th back-step }
It is desirable that when executing a small number of back-steps, their cost 
would be constant. More generally, we would like the cost of a back-step to be a 
function of the number of back-steps executed. That is, that the $i$'th back-step 
costs $O(\log i)$, Rather than $O(\log n)$. When starting back-steps after a 
sequence of $n$ forward steps, where $n$ is a power of $2$, using the above algorithm, 
then in fact we do already obtain a cost of at most $\log i$ per back-step. Indeed, the 
cost is proportional to the number of red-pebbles and this number is at most $\log i$ 
for the $i$'th back-step, since the largest tree affected when traversing $i$ 
steps backward is of size at most $2i$. However, using the above algorithm for 
arbitrary $n$, starting back traversal after moving forward $n$ steps could cost 
$\Theta(\log n)$ per back-step. 

We modify the implementation of back-step so as to obtain cost of $O(\log i)$ 
regardless of the point in which the back traversal begins, by creating red pebbles 
mostly in a tree whose size is at most $4i$. Creation of red pebbles in most 
of the higher nodes is delayed to future back-steps. 


\begin{Claim}{smallest-sub-tree}
Consider the smallest sub-tree $T$ that includes both nodes representing positions $n$
and $n-i$, and let $0 < j\le i$. (Note that this tree could be of size up to $2n$ even
for very small $i$). Then, the number of red-pebbles required for the node representing
$n-j$ within the tree is at most $\log i \over 2$$+1$.  
\end{Claim}

\begin{Proof}
Since the tree $T$ is the smallest that includes the sequence of nodes representing 
positions $n-j$, $0<j\le i$, then the right child of the root of this tree represents
a node $n-j'$ in the sequence, for some $0<j' \le i$. We consider separately the two 
parts of the sequence.

The suffix of the sequence, $n-j'-1,\ldots ,n-i$, begins at the right leaf of the 
left sub-tree of $T$ (the situation is similar to back-traversal from a position $n$ 
that is power of $2$), and is fully contained in a sub-tree $T'$ of size at most $2i$.
Within $T'$, the number of red pebbles is at most $\log i + 1 \over 2$, and no 
additional red pebble is required since the root of $T'$ is on a right path from the 
left child of the root of $T$. 

The prefix of the sequence, $n,\ldots,n-j'$, consists of two sub-sequences, one on a 
left path $p$ starting at the right child of the root of $T$, and the other fully 
contained in a sub-tree $T''$ whose root is on $p$, and whose size is at most $2i$.
One pebble is required for $p$, and within $T''$, the number of red pebbles is at 
most $\log i + 1 \over 2$, totaling at most $\log i +1 \over 2$$ + 1$.
\end{Proof}

The algorithm is as follows. Let $R$ be the smallest right sub-tree that fully 
contains the tree $T$ (as a sub-tree) from the claim. Then, at every back-step only 
red pebbles within $R$ are created or moved. Each pebble except for the highest 
pebble is moved two steps per every back-step, as before. The highest pebble is 
moved at up to double pace, that is up to 4 steps for every back-step. 
The double pace covers the delay this pebble has had, when it was not in the tree $R$ 
(when back traversing within the right sub-tree of $R$). Creation or move of 
red-pebbles outside $R$ is delayed until a later stage. 

>From the claim it follows that the number of red-pebbles touched at every back-step 
is at most $\log i + 1 \over 2$$ + 2$. Each pebble, except for the highest pebble,
is moved two steps for every back-step; the highest pebble is moved at most 4 steps 
for every back-step. The total number of steps per back-step is therefore at 
most $\log i + 7$.

It remains to show that the delay of high pebbles is properly compensated by the 
double pace approach. It suffices to show that the lowest delayed red pebble is in 
place when first needed. Let $L$ be the sub-tree whose root is the left sibling of 
the root of $R$. The size of $L$ is at least $i$, and once entering $L$, the lowest 
delayed red pebble starts moving in double pace. Therefore, when finishing the traversal 
through $L$, the green pebble created by that delayed pebble would already be in place. 
}
\RemoveICALP{
\section{Traversals of Trees and other graphs}

Next we discuss implementations for other linked structures. For every structure 
(tree, dag, graph), a (forward) traversal defines a linked list, which we call 
traversal list, from the starting point (``root'') to the current position. This 
traversal list consists of the sequence of nodes in order of traversal (possibly 
with repetitions in case of a general graph). A back-step from a node in the 
traversal list is always defined as getting to the preceding node in the traversal list.

Since the back-step is always implemented by first going to a certain pebble and 
then moving forward from the pebble, we will not be much concerned with having 
in-degree larger than one. Rather, we will need to provide attention to out-degree 
larger than one. We will therefore first consider directed trees. The extensions to 
dags and general graphs will be straightforward. 

\subsection{Trees}
Assume that we are given a directed tree, in which edges are directed from 
nodes to their children. A (forward) traversal on the tree is to move from a node 
to one of its children. A back-step from a given node is moving to its parent. 
Without additional information, a back-step can only be done by starting at the root 
and following the footsteps of the forward traversal (this can only be done under 
certain assumption to be discussed later). This requires only 2 pebbles, but requires 
time proportional to the depth of the current position.

Our objective is to support effective back-steps as well as back-queries, using small 
space. As for the case of list, a trivial support for back traversal is to add 
a back edge for every edge. A more effective approach is to add trailing back pointers 
as we go (that is, when positioned at node i we have a back path from i to the root).

Our solution for a directed list extends quite naturally to directed trees. 
We consider two types of trees:\\
{\bf Type A:} When positioned at a node v, then given the identity of a node $u$ 
that is a descendant of $v$, it is possible to determine in constant time which of 
the children of $u$ is an ancestor of $v$.\\  
{\bf Type B:}  When positioned at a node $v$, if $v$ has $r$ children, then 
%
%
$\log r$ bits can be used to identify each child in constant time 
(e.g., by taking the ranks in a lexicographic order of their names).

Note that the trivial solution for (expensive) back-steps using only 2 pebbles only 
works for a Type A tree. For a Type B tree, a simple correction of the algorithm is 
to maintain, for every node along the path from the root to the current position, 
the identifying information regarding which child should be traversed next. 
For a node $v$,  the number of bits required for this identifying information is 
$R(v) = \sum_{u\ Is\ an\ ancestor\ of\ v}{\log d(u)}$, where $d(u)$ is the number 
of children of $u$. For a binary tree, $R(v)$ is simply the depth of $v$.

\Paragraph{Traversals of Type A trees}

As we traverse forward in the tree, we maintain a data structure corresponding to 
the list defined by the traversal. Specifically, if $p$ is the path from the root to 
the current position, then the current data structure corresponds to a traversal 
along $p$. In the implementation of a back-step or of a back query, when moving forward 
from a pebble the identity of the child to which we need to move forward is determined 
by the property of Type A trees. 

The only possible complication that one could anticipate is when having a sequence 
of back-steps, followed by moving forward along a different path. For the implementation 
that guarantees $\log n$ cost per back-step, the data structure after returning 
to a node is similar to what it is arriving when arriving at the node in the first 
time, except for possibly pebbles that were added during the doubling phase, and are 
now redundant.

For the implementation that guarantees $\log i$ cost for the $i$'th back-step, 
the situation is a bit subtler, since the delay while backtracking should be 
accounted for. Fortunately, the delay only affects pebbles that are quite close 
to the root, and are not affected by the backtrack sequence. Specifically, the 
following shows that the implementation works here as well. 

\Paragraph{Traversals of Type B trees}

The algorithms for Type A trees works also for Type B trees, except that we need 
to resolve the identity issue as discussed above. By maintaining this information, 
using $R(v)$ additional bits when positioned at node $v$, we can indeed implement 
the algorithm described for Type A trees.

Comment: Any hierarchical structure that identifies nodes by their logical path 
is typically of type A. For instance: a file descriptor or a URL description. The 
``dot-dot'' operation is in fact going up a tree, and the Type A algorithm discussed 
above enables effective support of this operation even for implementations that only 
use unidirectional links.

\subsection{Directed Graphs}
We consider first directed acyclic graphs (DAGS). We only consider Type B DAGS 
(with the natural extension from trees). The implementation of back-steps in Type B 
trees does not rely on the fact that the in-degree of nodes is one (since it is 
implemented by moving forward on the tree). The very same algorithm works for DAGs 
as well. The existences of cycles have no effect on the algorithm.
}
}

\begin{figure*}
\centering
\epsfig{file=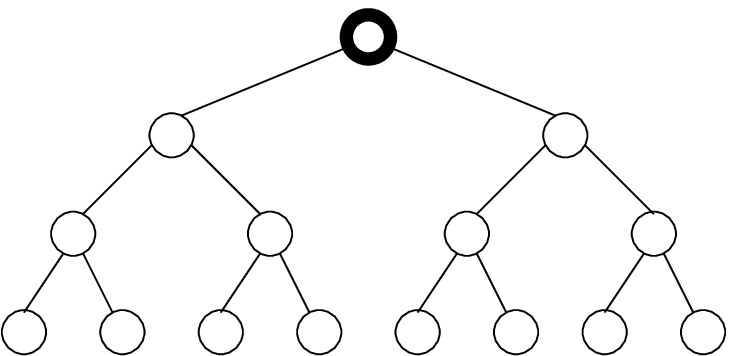,height=1.1in,width=1.7in,}
\quad
\epsfig{file=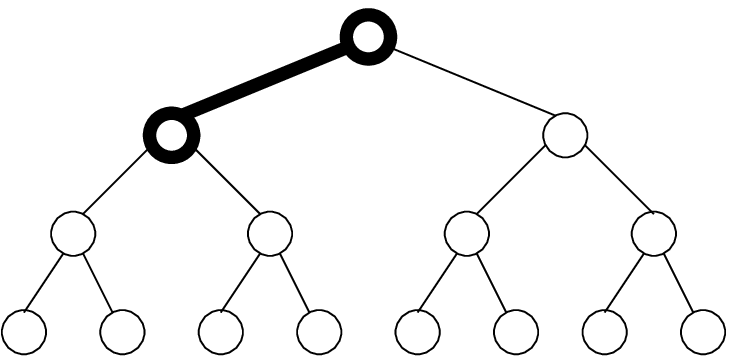,height=1.1in,width=1.7in,}
\quad
\epsfig{file=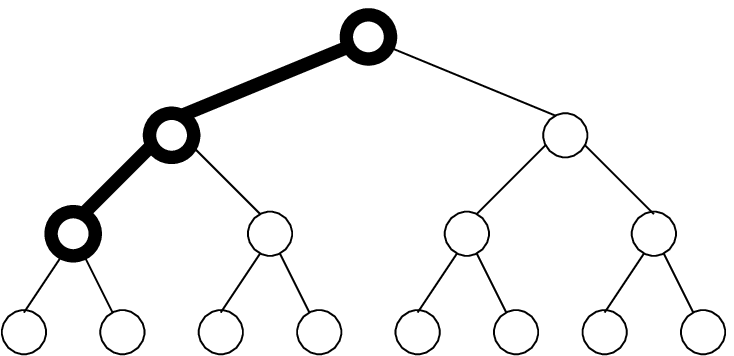,height=1.1in,width=1.7in,}
\quad
\epsfig{file=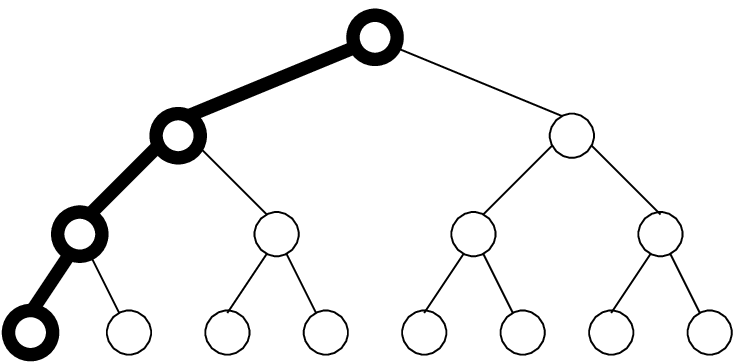,height=1.1in,width=1.7in,}
\quad
\epsfig{file=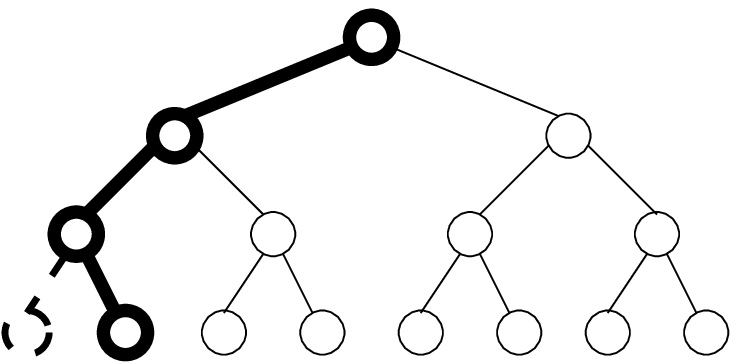,height=1.1in,width=1.7in,}
\quad
\epsfig{file=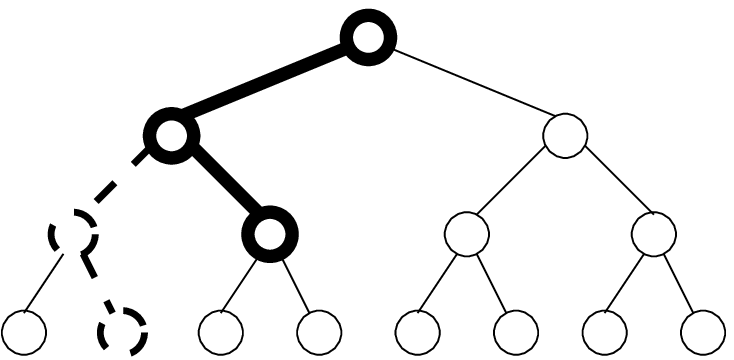,height=1.1in,width=1.7in,}
\quad
\epsfig{file=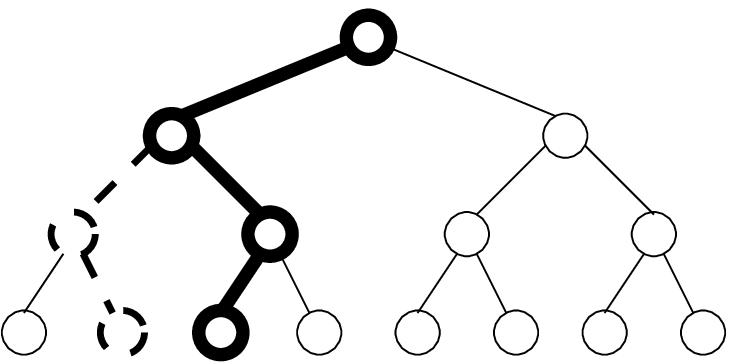,height=1.1in,width=1.7in,}
\quad
\epsfig{file=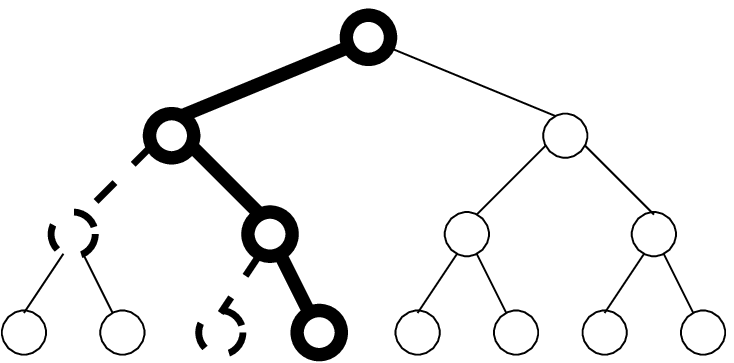,height=1.1in,width=1.7in,}
\quad
\epsfig{file=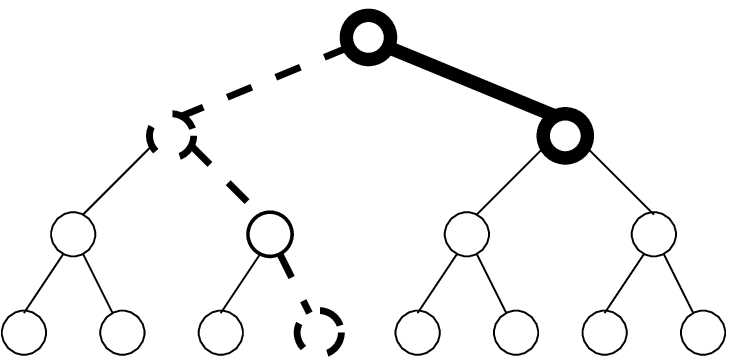,height=1.1in,width=1.7in,}
\caption{The list traversal synopses in 9 steps of forward
traversal from the beginning of the list 
(left to right, top to bottom). 
Nodes in bold face are blue pebbled, and bold
edges constitute the blue paths. Fragmented nodes are green pebbled, 
and fragmented  edges constitute green paths.
}
\label{forward-traversal}
\end{figure*}
\bibliographystyle{abbrv}
\bibliography{paper}
\end{document}

%% file: macros.tex
\def\NewSymbol#1[#2]/*#3*/{%
	\gdef#1{#2%
	}%
}%

\def\BibTeX{{\rm B\kern-.05em{\sc i\kern-.025em b}\kern-.08em
    T\kern-.1667em\lower.7ex\hbox{E}\kern-.125emX}}
\renewcommand{\baselinestretch}{\LineSpace}

\newcommand{\mylarge}{\large\renewcommand{\baselinestretch}{\LineSpace}}

\def\SizeExpectancy[#1]#2{{{\bf E}\mathopen#1({#2}\mathclose#1)}}

\def\SizeSquareExpectancy[#1]#2{{{\bf E}^2\mathopen#1({#2}\mathclose#1)}}

\def\SizeVariance[#1]#2{{{\bf Var}\mathopen#1({#2}\mathclose#1)}}

\def\SizeProbability[#1]#2{{{\bf Prob}\mathopen#1({#2}\mathclose#1)}}

\def\Log(#1){{\log \left({#1}\right)}}
\def\Ln(#1){{\ln \left({#1}\right)}}

\def\Range[#1,#2]{[#1,#2]}
\def\SetRange[#1,#2]{\left\{#1,\ldots{},#2\right\}}
\def\Binomial[#1,#2]{{{#1}\choose{#2}}}
\def\StirlingFirst[#1,#2]{{{#1}\brack{#2}}}
\def\StirlingSecond[#1,#2]{{{#1}\brace{#2}}}
\def\Euler[#1,#2]{{{#1}\atopwithdelims<>{#2}}}
\def\Atmost(#1){{O\!\left({#1}\right)}}
\def\Atleast(#1){{\Omega\!\left({#1}\right)}}
\def\SizeParenthesis[#1]#2{{\mathopen#1({#2}\mathclose#1)}}

\newenvironment{Theorem}[1]{
    \theorem
    \protect\label{Theorem:#1}
	\sloppypar
}{
	\endsloppypar
    \endtheorem
}
\newenvironment{Named:Theorem}[2]{
	\sloppypar
    \theorem[{#2}]
    \protect\label{Theorem:#1}
}{
    \endtheorem
	\endsloppypar
}
\newenvironment{Lemma}[1]{
	\sloppypar
    \lemma
    \protect\label{Lemma:#1}
}{
    \endlemma
	\endsloppypar
}
\newenvironment{Named:Lemma}[2]{
	\sloppypar
    \lemma[{#2}]
    \protect\label{Lemma:#1}
}{
    \endlemma
	\endsloppypar
}

\newenvironment{Named:Corollary}[2]{
	\sloppypar
    \corollary[{#2}]
    \protect\label{Corollary:#1}
}{
    \endcorollary
	\endsloppypar
}

\newenvironment{Named:Proposition}[2]{
	\sloppypar
    \proposition[{#2}]
    \protect\label{Proposition:#1}
}{
    \endproposition
	\endsloppypar
}
\newenvironment{Claim}[1]{
    \claim
    \protect\label{Claim:#1}
	\sloppypar
}{
	\endsloppypar
    \endclaim
}
\newenvironment{Named:Claim}[2]{
	\sloppypar
    \claim[{#2}]
    \protect\label{Claim:#1}
}{
    \endclaim
	\endsloppypar
}

\newenvironment{Named:Definition}[2]{
    \pagebreak[1]
    \definition[{#1}]
    \protect\label{Definition:#2}
	\sloppypar
}{
	\endsloppypar
    \enddefinition
    \pagebreak[1]
}

\newenvironment{Named:Fact}[2]{
    \pagebreak[1]
    \fact[{#1}]
    \protect\label{Fact:#2}
	\sloppypar
}{
	\endsloppypar
    \endfact
    \pagebreak[1]
}

	\newenvironment{Design Recipe}[1]{
	\sloppypar
	\bigskip
	\begin{quote}
	\design
	\sf
    \protect\label{Design Recipe:#1}
}{
    \enddesign
	\end{quote}
	\bigskip
	\endsloppypar
}

\renewenvironment{Design Recipe}[1]{
    \design
    \protect\label{Design Recipe:#1}
	\sloppypar
}{
	\endsloppypar
    \enddesign
}
\newenvironment{Named:Design Recipe}[2]{
    \pagebreak[1]
	\bigskip
	\begin{quote}
    \design
	\sf
    \protect\label{Design Recipe:#1}
	{\rm(#2)}
	\sloppypar
}{
	\endsloppypar
    \enddesign
	\end{quote}
	\bigskip
    \pagebreak[1]
}

\renewenvironment{Named:Design Recipe}[2]{
\begin{recipe}[H]
	\begin{center}
	\begin{minipage}{0.90\textwidth}
	\hline
	\sf
    \protect\label{Design Recipe:#1}
	\caption{#2}
}{
	\vspace{2ex}
	\hline
	\end{minipage}
	\end{center}
\end{recipe}
}

\renewenvironment{Named:Design Recipe}[2]{
	\sloppypar
    \design[{#2}]
    \protect\label{Design Recipe:#1}
}{
    \enddesign
	\endsloppypar
}

\renewenvironment{Named:Design Recipe}[2]{
\begin{recipe}[H]
	\begin{center}
	\begin{minipage}{0.90\textwidth}
	\it
    \protect\label{Design Recipe:#1}
	{\protect\caption{\bf #2}}
}{
	\end{minipage}
	\end{center}
\end{recipe}
}

\newcommand{\Ref}[2]{#1~\ref{#1:#2}}
\def\squareforqed{\rule[0.4ex]{0.8ex}{0.8ex}}
\def\qed{\ifmmode\squareforqed\else{\unskip\nobreak\hfil
\penalty50\hskip1em\null\nobreak\hfil\squareforqed
\parfillskip=0pt\finalhyphendemerits=0\endgraf}\fi}
\newenvironment{Proof}{\paragraph{{\em Proof. }}}{%
\nobreak\hfill\qed\par
	\pagebreak[1]
}
\newenvironment{Proof:Of}[2]{\paragraph{Proof of \Ref{#1}{#2}}
}{
}
\newcommand{\Em}[1]{{\em #1\/}}

\newcommand{\TextComment}[1]{}

\newcommand{\Section}[2]{\section{#1}\label{Section:#2}}
\newcommand{\Subsection}[2]{\subsection{#1}\label{Section:#2}}

\newcommand{\Paragraph}[1]{\noindent{\bf #1.\ }}

\newcommand{\SmallItems}{%
\setlength{\parskip}{0in}%
\setlength{\topsep}{0in}%
\setlength{\itemsep}{0in}%
\setlength{\parsep}{0in}%
\setlength{\labelwidth}{0in}%
}
\newcommand{\OneSmallItem}{%
\setlength{\itemsep}{0in}%
\setlength{\parsep}{0in}%
\setlength\labelsep{1ex}%
\setlength{\itemindent}{2ex}%
\setlength{\rightmargin}{0in}\setlength{\leftmargin}{0ex}%
}

\newcounter{AlphaCount}

\newcounter{RomaniaCount}

\def\PageLabel[#1]{\label{page:#1}}

\let\log=\lg

\def\StirlingTwo(#1,#2){{{#1} \brace {#2}}}